\documentclass[12pt]{article}
\usepackage{amsmath, amssymb, amsfonts}
\usepackage{graphicx}
\usepackage{hyperref}
\usepackage{geometry}
\usepackage{caption}
\usepackage{authblk}
\usepackage{float}
\usepackage{multirow}
\usepackage{booktabs}
\usepackage{enumitem}
\usepackage{amsmath}
\title{\textbf{QESM: A Leap Towards Quantum-Enhanced ML Emulation Framework for Earth and Climate Modeling}}

\author[1]{Adib Bazgir}
\author[1]{Yuwen Zhang\thanks{Corresponding Email Address: zhangyu@missouri.edu}}

\affil[1]{Mechanical and Aerospace Engineering Department\\ University of Missouri-Columbia\\ Missouri, USA}

\begin{document}

\maketitle
\begin{abstract}
Current climate models often struggle with accuracy because they lack sufficient resolution, a limitation caused by computational constraints. This reduces the precision of weather forecasts and long-term climate predictions. To address this issue, we explored the use of quantum computing to enhance traditional machine learning (ML) models. We replaced conventional models like Convolutional Neural Networks (CNN), Multilayer Perceptrons (MLP), and Encoder-Decoder frameworks with their quantum versions: Quantum Convolutional Neural Networks (QCNN), Quantum Multilayer Perceptrons (QMLP), and Quantum Encoder-Decoders (QED). These quantum models proved to be more accurate in predicting climate-related outcomes compared to their classical counterparts. Using the ClimSim dataset, a large collection of climate data created specifically for ML-based climate prediction, we trained and tested these quantum models. Individually, the quantum models performed better, but their performance was further improved when we combined them using a "meta-ensemble" approach, which merged the strengths of each model to achieve the highest accuracy overall. This study demonstrates that quantum machine learning can significantly improve the resolution and accuracy of climate simulations. The results offer new possibilities for better predicting climate trends and weather events, which could have important implications for both scientific understanding and policy-making in the face of global climate challenges.\end{abstract}
\vspace{20pt}

\textit{\textbf{Keywords}} Quantum Machine Learning . Climate and Earth Modeling . Long-term Prediction . Meta-ensemble Approach . CNN . MLP . Encoder-Decoder (ED)

\newpage
\section{Introduction}
Climate change policy heavily relies on predictions from numerical physical simulations, yet these simulations often struggle with accurate representations of cloud physics and extreme rainfall events \cite{Change2007}. This deficiency persists despite leveraging the capabilities of advanced supercomputers, primarily due to the complex nature of Earth system interactions that demand significant compromises in spatial resolution. Traditional climate simulations use empirical mathematical "parameterizations" to represent sub-resolution physical processes, but these are fraught with assumptions that introduce errors, potentially distorting future climate predictions. Machine learning (ML) presents a promising solution by emulating these complex sub-resolution processes at reduced computational complexities, potentially enhancing both the cost-effectiveness and accuracy of climate simulations \cite{Schneider et al.2017}. Hybrid-ML climate simulators, which combine traditional numerical methods with ML-based emulators, offer a novel approach to overcome the resolution limitations of existing simulators. These emulators are trained on data from high-resolution simulations to predict large-scale atmospheric changes resulting from unresolved small-scale physics, aiming to replace heuristic assumptions with data-driven insights. Despite the conceptual advances, operational implementation of hybrid-ML simulators remains nascent, hindered by the scarcity of suitable training data. To address this, multi-scale simulation methods have been proposed to generate comprehensive training datasets that integrate seamlessly with coarse-resolution simulators. The introduction of "ClimSim" \cite{Yu et al.2024}, the most extensive dataset designed for training ML emulators for atmospheric phenomena, marks a significant step forward. This dataset is intended to lower entry barriers for ML practitioners and foster the development of robust frameworks to improve the accuracy and performance of climate models, ultimately aiding long-term climate projections.

Recent research has explored the development of hybrid machine learning (ML) models for simulating climate systems, similar to the ClimSim approach \cite{Gentine et al.2018}. While many studies have focused on simplified models, often with limited geographical representation \cite{Lin et al.2023,Wang et al.2022} or neglecting essential land-surface variables, the ClimSim dataset stands out for its extensive coverage of relevant variables across various scenarios (e.g., aquaplanet vs. real geography) \cite{Gentine et al.2018}. Most prior work utilized simpler ML architectures, whereas ClimSim's complexity might necessitate more advanced models like ResNet or variational encoder-decoders \cite{Behrens et al.2022}. Though real-world applications of these hybrid models remain challenging, some promising results have emerged, suggesting potential for enhanced accuracy and stability. Further, methods have been proven to enforce physical constraints \cite{Beucler et al.2021,Reed et al.2023}. ClimSim's comprehensive data could inspire innovative ML techniques that could directly benefit existing climate models currently used by leading organizations such as the U.S. Department of Energy \cite{Gentine et al.2018}. Beyond multi-scale modeling, significant research has leveraged similar hybrid machine learning methods to improve the accuracy of uniform resolution climate models. This includes advancements in operational models with land coupling, enhancing their stability. Further research explores full model emulation (FME) for short-term weather forecasting. However, applying this approach to climate modeling using high-frequency output data remains a challenge. Notably, recent work indicates that incorporating spherical geometry and resolution invariance via spherical Fourier neural operators can improve long-term prediction stability \cite{Pathak et al.2022}. While ClimSim enjoys crucial advantages of hybrid-ML climate simulation, comprehensive sampling of atmospheric state variables, and advanced baseline models, our proposed integrated QESM model, shown in Figure 1, containing QCNN, QMLP, and QED further outperforms the existing ClimSim model in term of the prediction accuracy. 

\section{Our Method: QESM}
\subsection{Dataset Generation}
As provided in Ref. \cite{Yu et al.2024}, ClimSim addresses a regression problem by mapping a multivariate input vector 
\((x \in \mathbb{R}^{d_i})\) of size \(d_i = 124\) and targets \((y \in \mathbb{R}^{d_0})\) of size \(d_0 = 128\). The input represents the local vertical structure of macro-scale state variables in a multi-scale physical climate simulator, including surface pressure, insolation, and latent/sensible heat flux. The target vector includes NETSW, FLWDS, PRECSC, PRECC, SOLS, SOLL, SOLSD, and SOLLD which are given in Table 1 and input/target study section. Due to the large volume of ClimSim datasets, 41.2TB and 1.488TB for high and low-resolution datasets, and a massive 9,800 GPU-hours usage for the training process, only a portion of the dataset (\(d_i^{\text{NEW}} = 4\) and \(d_0^{\text{NEW}} = 8\)) is utilized for a testament in this study.
\begin{figure}[H]
    \centering
    \includegraphics[width=0.95\textwidth]{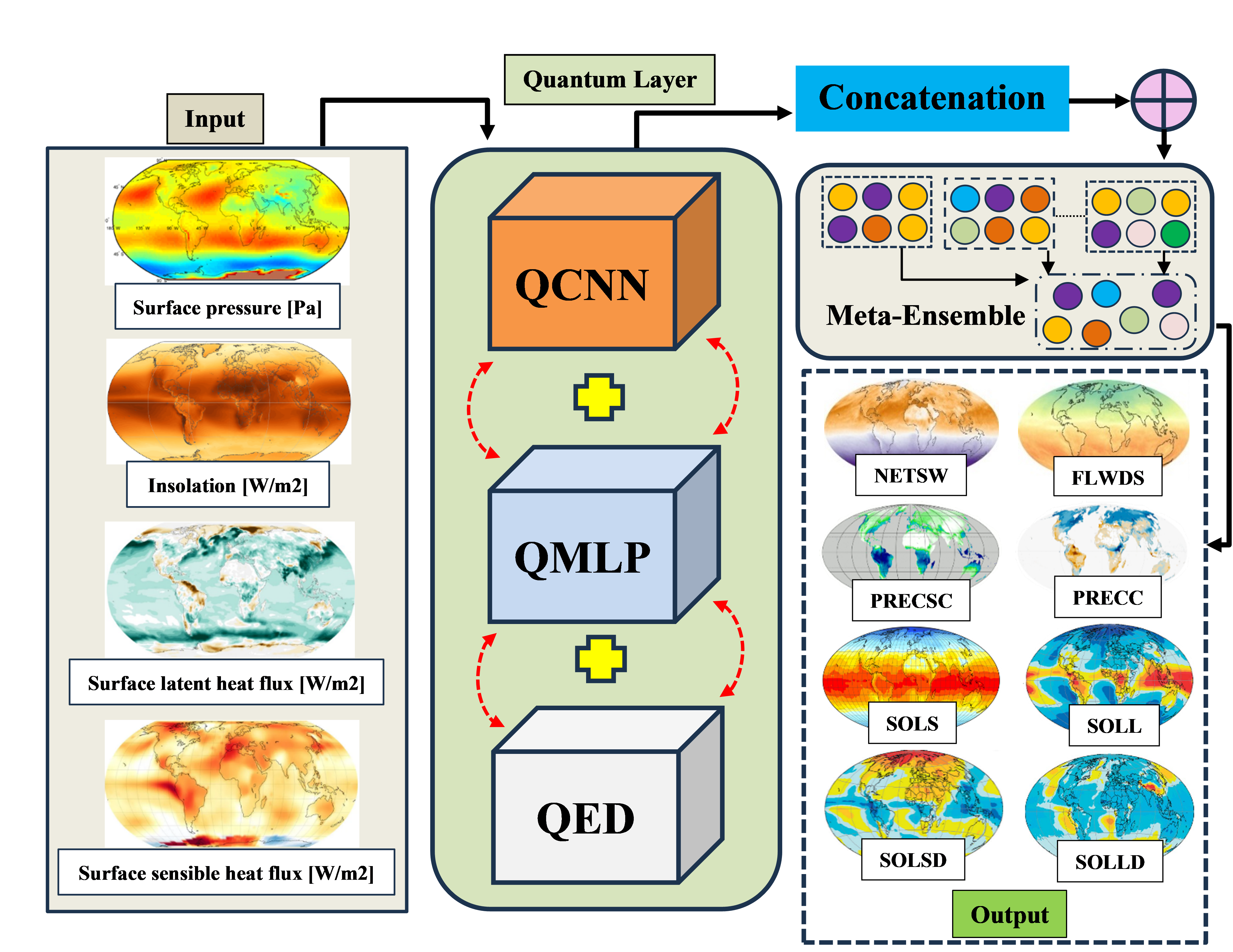}
    \caption{The overall schematic of QESM model equipped with QCNN, QMLP, QED, and Meta-ensemble approach.}
    \label{fig:qed}
\end{figure}

\subsection{Input and Target Study}
More details on the target vector and its components are further disclosed as following:

\begin{itemize}[left=0pt]
\item \textbf{Surface Pressure [Pa]}

Surface pressure is the atmospheric pressure at a specific location on the Earth's surface, measured in Pascals (Pa). It represents the force per unit area exerted by the weight of the air column above that location. Surface pressure is fundamental in meteorology and climatology, influencing weather patterns and climatic conditions. Variations in surface pressure drive winds and affect precipitation, impacting human activities and engineering applications such as aviation and construction. The hydrostatic equation describes the change in pressure with altitude in the atmosphere:
\begin{equation}
\frac{dP}{dz} = -\rho g
\end{equation}
where \( P \) is the atmospheric pressure, \( z \) is the altitude, \( \rho \) is the air density, and \( g \) is the acceleration due to gravity. At the surface, the pressure is the integral of the air density and gravitational force:
\begin{equation}
P = \int_0^z \rho g \, dz
\end{equation}
    
\item \textbf{Insolation [W/m$^2$]}
    
Insolation is the amount of solar radiation energy received per unit area on a surface during a given time, typically measured in watts per square meter (W/m$^2$). It is critical in understanding Earth's energy balance, climate modeling, and solar energy applications. Insolation affects surface temperature, weather patterns, and the efficiency of solar power systems. Insolation can be calculated using:
\begin{equation}
I = S(1 - \alpha) \cos(\theta)
\end{equation}
where \( I \) is the insolation, \( S \) is the solar constant (approximately 1361 W/m$^2$ at the top of the atmosphere), \( \alpha \) is the surface albedo (reflectivity), and \( \theta \) is the angle of incidence of the sunlight.

\item \textbf{Surface Latent Heat Flux [W/m$^2$]}
    
Surface latent heat flux is the energy per unit area transferred from the Earth's surface to the atmosphere due to the phase change of water, such as evaporation or condensation, measured in watts per square meter (W/m$^2$). Latent heat flux is a crucial component of the hydrological cycle and Earth's energy budget, influencing weather patterns, climate, and atmospheric moisture distribution. Latent heat flux is given by:
\begin{equation}
Q_l = \lambda E
\end{equation}
where \( Q_l \) is the latent heat flux, \( \lambda \) is the latent heat of vaporization (approximately 2.5 $\times$ 10$^6$ J/kg for water), and \( E \) is the evaporation rate (kg/m$^2$/s). The evaporation rate can be estimated using the Penman-Monteith equation.
    
\item \textbf{Surface Sensible Heat Flux [W/m$^2$]}
    
Surface sensible heat flux is the heat energy transferred between the Earth's surface and the atmosphere due to temperature differences, expressed in watts per square meter (W/m$^2$). Sensible heat flux affects temperature variations and atmospheric stability, and it is important for weather forecasting, climate studies, and understanding energy exchanges between the surface and the atmosphere. Sensible heat flux can be expressed as:
\begin{equation}
Q_s = \rho c_p \frac{dT}{dz}
\end{equation}
where \( Q_s \) is the sensible heat flux, \( \rho \) is the air density, \( c_p \) is the specific heat capacity of air at constant pressure, and \( \frac{dT}{dz} \) is the temperature gradient near the surface. A practical form using the bulk aerodynamic formula is:
\begin{equation}
Q_s = \rho c_p C_h U (T_s - T_a)
\end{equation}
where \( C_h \) is the heat transfer coefficient, \( U \) is the wind speed, \( T_s \) is the surface temperature, and \( T_a \) is the air temperature near the surface.

\item \textbf{Net Surface Shortwave Flux (NETSW) [W/m$^2$]}
    
Net surface shortwave flux (NETSW) is the balance between incoming and reflected shortwave solar radiation at the Earth's surface, measured in watts per square meter (W/m$^2$). It is a critical component of the Earth's surface energy budget, influencing surface temperature and climate. NETSW accounts for the portion of solar radiation that is absorbed by the surface after subtracting the reflected part.
\begin{equation}
\text{NETSW} = (1 - \alpha) \times S_{\text{down}}
\end{equation}
where \( \alpha \) is the surface albedo (reflectivity), and \( S_{\text{down}} \) is the incoming shortwave solar radiation.
    
\item \textbf{Downward Surface Longwave Flux (FLWDS) [W/m$^2$]}
    
Downward surface longwave flux (FLWDS) is the amount of longwave (infrared) radiation emitted from the atmosphere that reaches the Earth's surface, measured in watts per square meter (W/m$^2$). This flux contributes to the warming of the surface and is a key component in the surface energy budget and greenhouse effect.
\begin{equation}
\text{FLWDS} = \varepsilon \sigma T_a^4
\end{equation}
where \( \varepsilon \) is the emissivity of the atmosphere, \( \sigma \) is the Stefan-Boltzmann constant (5.67 $\times$ 10$^{-8}$ W/m$^2$K$^4$), and \( T_a \) is the temperature of the atmosphere.
    
\item \textbf{Snow Rate (PRECSC) [m/s]}
    
Snow rate (PRECSC) is the rate at which snow precipitates from the atmosphere to the surface, measured in meters per second (m/s). This parameter is crucial for understanding snowfall accumulation, hydrology, and climate dynamics in snow-covered regions. The snow rate can be measured directly by precipitation gauges or inferred from weather radar and satellite data, typically expressed as:
\begin{equation}
\text{PRECSC} = \frac{dH_{\text{snow}}}{dt}
\end{equation}
where \( H_{\text{snow}} \) is the height of snow accumulation, and \( t \) is time.

\item \textbf{Rain Rate (PRECC) [m/s]}
    
Rain rate (PRECC) is the rate at which rain precipitates from the atmosphere to the surface, measured in meters per second (m/s). It is vital for hydrological studies, weather forecasting, and climate modeling, affecting water resources and ecosystems. The rain rate can be measured by rain gauges or inferred from weather radar and satellite data, typically expressed as:
\begin{equation}
\text{PRECC} = \frac{dH_{\text{rain}}}{dt}
\end{equation}
where \( H_{\text{rain}} \) is the height of rain accumulation, and \( t \) is time.
    
\item \textbf{Visible Direct Solar Flux (SOLS) [W/m$^2$]}
    
Visible direct solar flux (SOLS) is the amount of solar radiation in the visible spectrum that directly reaches the Earth's surface, measured in watts per square meter (W/m$^2$). This parameter is crucial for understanding solar energy input, photosynthesis, and surface heating. The SOLS parameter can be determined by:
\begin{equation}
\text{SOLS} = S_0 \cos(\theta) e^{-\tau}
\end{equation}
Where \( S_0 \) is the solar constant, \( \theta \) is the solar zenith angle, and \( \tau \) is the atmospheric optical depth.
    
\item \textbf{Near-IR Direct Solar Flux (SOLL) [W/m$^2$]}
    
Near-IR direct solar flux (SOLL) is the amount of solar radiation in the near-infrared spectrum that directly reaches the Earth's surface, measured in watts per square meter (W/m$^2$). This flux is important for understanding surface heating and energy balance, particularly in arid and semi-arid regions.
\begin{equation}
\text{SOLL} = S_{\text{IR}} \cos(\theta) e^{-\tau}
\end{equation}
where \( S_{\text{IR}} \) is the near-infrared portion of the solar constant, \( \theta \) is the solar zenith angle, and \( \tau \) is the atmospheric optical depth.

\item \textbf{Visible Diffused Solar Flux (SOLSD) [W/m$^2$]}
    
Visible diffused solar flux (SOLSD) is the amount of solar radiation in the visible spectrum that reaches the Earth's surface after being scattered by the atmosphere, measured in watts per square meter (W/m$^2$). It contributes to overall solar energy input and is significant for solar energy applications and ecological processes.
\begin{equation}
\text{SOLSD} = S_0 (1 - e^{-\tau}) \cos(\theta)
\end{equation}
where \( S_0 \) is the solar constant, \( \tau \) is the atmospheric optical depth, and \( \theta \) is the solar zenith angle.
    
\item \textbf{Near-IR Diffused Solar Flux (SOLLD) [W/m$^2$]}
    
Near-IR diffused solar flux (SOLLD) is the amount of solar radiation in the near-infrared spectrum that reaches the Earth's surface after being scattered by the atmosphere, measured in watts per square meter (W/m$^2$). This flux influences surface energy balance and heating, especially in regions with high aerosol concentrations.
\begin{equation}
\text{SOLLD} = S_{\text{IR}} (1 - e^{-\tau}) \cos(\theta)
\end{equation}
where \( S_{\text{IR}} \) is the near-infrared portion of the solar constant, \( \tau \) is the atmospheric optical depth, and \( \theta \) is the solar zenith angle.
    
\end{itemize}

\subsection{Preprocessing Model Workflow}
To guide machine learning (ML) practitioners using ClimSim, an example workflow is provided for a low-resolution, real-geography dataset. The task involves emulating a subset of total input and target variables (4 inputs and 8 targets), similar to recent literature \cite{Yu et al.2024}. The dataset is split into an 8-year training/validation set, with the first 7 years used for training and the last year for validation. For each sample, horizontal location and time are collapsed into a single sample dimension. Variables are normalized by subtracting the mean and dividing by the range, calculated separately for each of the input and target variables. Finally, variables are concatenated into multivariate input and output vectors for each sample. 

\begin{table}[h]
    \centering
    \caption{Input and target details of the QESM model.}
    \begin{tabular}{llcc}
        \toprule
        \textbf{Input $[d_i^{\text{NEW}}]$} & \textbf{Size} & \textbf{Target $[d_0^{\text{NEW}}]$} & \textbf{Size} \\
        \midrule
        Surface pressure [Pa] & 1 & NETSW [W/m$^2$] & 1 \\
        Insolation [W/m$^2$] & 1 & FLWDS [W/m$^2$] & 1 \\
        Surface latent heat flux [W/m$^2$] & 1 & PRECSC [m/s] & 1 \\
        Surface sensible heat flux [W/m$^2$] & 1 & PRECC [m/s] & 1 \\
        \multirow{4}{*}{} & & SOLS [W/m$^2$] & 1 \\
        & & SOLL [W/m$^2$] & 1 \\
        & & SOLSD [W/m$^2$] & 1 \\
        & & SOLLD [W/m$^2$] & 1 \\
        \bottomrule
    \end{tabular}
\end{table}

\subsection{Quantum Model Architectures}
Three quantum baseline models are comprehensively discussed along this study where more detailed schematics are provided in subsequent sections.

\subsubsection{Quantum Convolutional Neural Network (QCNN)}
The QCNN model architecture is represented in Figure 2 for processing spatial data through quantum convolutional and neural network layers to produce multiple outputs. The input data, a map-like earth distribution, is divided into smaller patches, each processed by Quantum Convolutional (QCONV) layers using quantum circuits. The QCONV 1D layers extract features by applying quantum gates to these patches. After the quantum convolutional layers, a quantum ReLU activation function (Q-RELU) introduces non-linearity. The processed patches then pass through a variational ansatz, a quantum circuit with trainable parameters that captures complex interactions. To prevent overfitting, quantum dropout layers (Q-Dropout) randomly drop certain quantum states during training. The model includes various quantum gates, such as Hadamard, RZ, RX, and RY, to further transform the data. These gates manipulate quantum states, allowing the network to learn complex data patterns. The final output consists of predicted classes, labeled as NETSW, FLWDS, PRECSC, PRECC, SOLS, SOLL, SOLSD, and SOLLD, representing different model predictions.

As shown in Figure 2, the QCNN architecture leverages the quantum computational framework to enhance the processing of complex climate data. The QCNN comprises several layers, each performing specific operations on the input data. The input to the QCNN is a multispectral image of the Earth, capturing various climate parameters. This image undergoes preprocessing, such as normalization and dimension adjustments, to ensure compatibility with the quantum convolutional layers. The first set of layers consists of Quantum Convolutional 1D (QCONV 1D) operations. The first QCONV 1D layer applies a quantum convolutional operation to the input data, extracting basic features from the multispectral image. This layer uses 32 filters with a kernel size of 3x3, a stride of 1, and same padding. The second QCONV 1D layer further refines features extracted by the first layer, capturing more complex patterns. This layer employs 64 filters with the same kernel size, stride, and padding as the first layer. Following the QCONV 1D layers, Quantum ReLU (Q-ReLU) activation functions are applied. The first Q-ReLU activation introduces non-linearity after the first QCONV 1D layer, enabling the network to learn complex data representations. The second Q-ReLU activation follows the second QCONV 1D layer, further enhancing the network’s capability to model intricate patterns. Quantum Dropout (Q-Dropout) layers are incorporated to prevent overfitting. The first Q-Dropout layer randomly deactivates 20\% of quantum gates during training, ensuring the network generalizes well to unseen data. The second Q-Dropout layer similarly deactivates 20\% of quantum gates in deeper layers, reducing overfitting risks.

The core quantum computational block includes a variational ansatz, a parameterized quantum circuit designed to approximate the desired quantum state transformations. This block features specific quantum gates such as Rx, Ry, and Rz, which apply rotations around the respective axes on the Bloch sphere. Entanglement operations, depicted by CNOT gates, create correlations between qubits, enhancing the network's capacity to model complex dependencies. The parameters within the variational ansatz are optimized during training to minimize the loss function. The final layer integrates the processed information, producing output predictions for various climate parameters. The architecture supports different climate-related outputs, such as NETSW, FLWDS, PRECSC, PRECC, SOLS, SOLL, SOLSD, and SOLLD. These outputs are essential for comprehensive climate analysis, providing insights into surface temperatures, precipitation patterns, and other critical metrics. The QCNN architecture harnesses the power of quantum computing to efficiently process high-dimensional climate data. Its ability to capture intricate patterns and dependencies within the data makes it a powerful tool for climate analysis, offering improved accuracy and efficiency compared to classical convolutional neural networks. Potential applications include climate modeling, weather forecasting, and environmental monitoring, where precise and timely predictions are crucial \cite{Kerenidis et al.2019}.

\begin{figure}[H]
    \centering
    \includegraphics[width=1\textwidth]{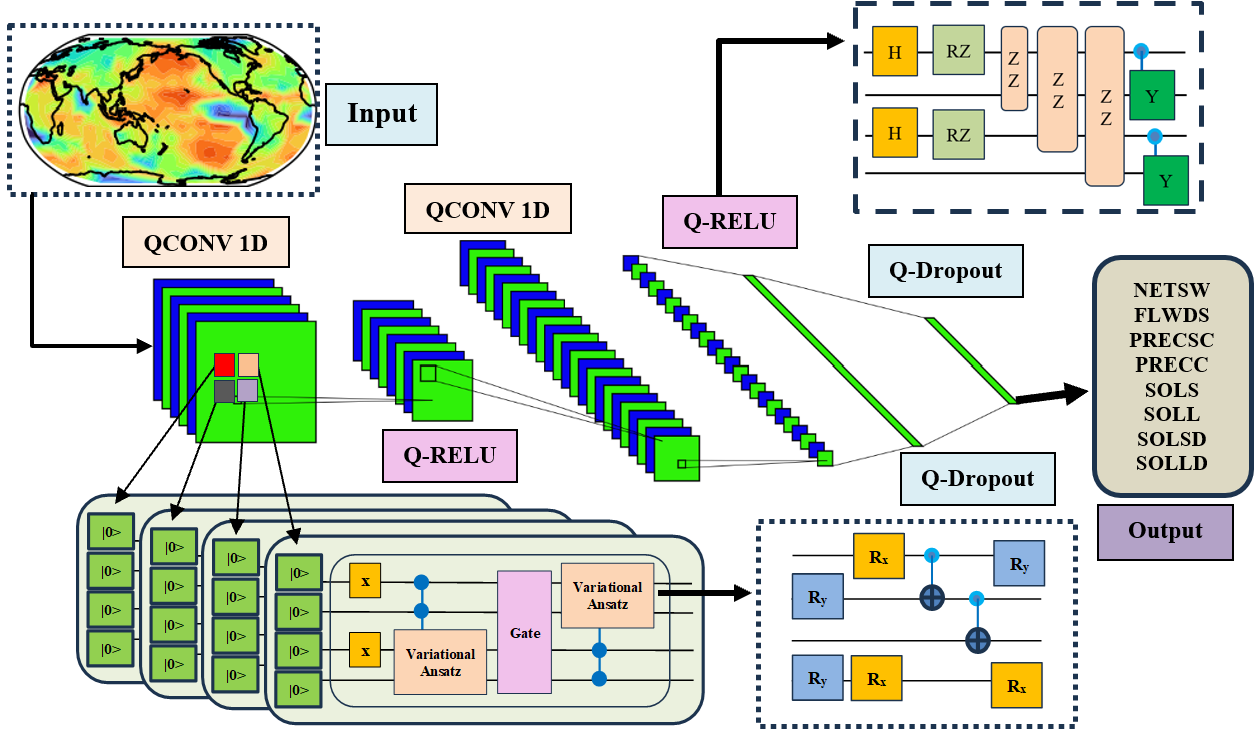}
    \caption{The detailed schematic of the QCNN model along with its quantum components.}
    \label{fig:qed}
\end{figure}

More specifically, the QCNN layer’s key steps and components involved in the quantum forward and backward passes for a QCNN layer are summarized as below, and the detailed discussion is given in Algorithm 1:

\begin{enumerate}[left=0pt]
\item \textbf{Quantum Convolution Product}:
    
The quantum convolution product is the primary operation, mapping the convolution process from classical CNNs to a quantum framework. It uses a mapping between the convolution of tensors and matrix multiplication, which can be reduced to inner product estimation between vectors.
    
    \item \textbf{Inner Product Estimation}:
    
    This involves calculating the inner product between the input and kernel tensors using quantum states. The inner product is estimated using amplitude estimation and median evaluation algorithms to ensure accuracy.
    
    \item \textbf{Non-Linearity}:
    
    After the convolution, a non-linear activation function (e.g., Q-ReLU) is applied. This is implemented using quantum circuits to handle the non-linearity, which is essential for the learning capability of neural networks.
    
    \item \textbf{Quantum Sampling}:
    
    The output of the quantum convolution is a quantum state representing the result of the convolution product. To retrieve meaningful classical information, quantum sampling techniques are used. This involves conditional rotations and amplitude amplification to enhance the probability of measuring important data points.
    
    \item \textbf{Quantum Tomography}:
    
    To convert the quantum state back to a classical form, quantum tomography with $\ell_\infty$ norm guarantees is employed. This process ensures that the classical output closely approximates the desired results from the quantum state.
    
    \item \textbf{Pooling Operation}:
    
    The pooling operation, which reduces the dimensionality of the data, is integrated into the QCNN structure. This can be performed during the QRAM update phase and includes techniques like maximum pooling or average pooling.

\end{enumerate}
\vspace{10pt}

\begin{table}[H]
    \centering
    \renewcommand{\arraystretch}{1.5}
    \begin{tabular}{p{0.95\textwidth}}
    \hline
    \textbf{Algorithm 1 (Combined Algorithm): QCNN with Quantum Backpropagation} \\
    \hline
    \textbf{Input:} Data input matrix $A^l$, kernel matrix $F^l$, precision parameters $\epsilon$, $\eta$, and $\delta$, non-linearity function $f$, learning rate $\lambda$. 
    
    \textbf{Output:} Updated data matrices $A^{l+1}$ and kernel matrices $F^l$. \\ \hline
    
    \textbf{1. Forward Pass (Quantum Convolution):} \\ 
    \textbf{• Inner Product Estimation:} \[
    \frac{1}{K} \sum_{p,q} \langle p | q \rangle \longrightarrow \frac{1}{K} \sum_{p,q} \langle p | q \rangle \langle \bar{P}_{pq} | g_{pq} \rangle
    \]
    
    \textbf{• Non-Linearity:} \[
    \frac{1}{K} \sum_{p,q} \langle p | q \rangle \left| f\left(Y_{p,q}^{l+1}\right) \right| |g_{pq}\rangle
    \]
    
    \textbf{• Quantum Sampling:} \[
    \frac{1}{K} \sum_{p,q} \alpha_{pq} \langle p | q \rangle \left| f\left(Y_{p,q}^{l+1}\right) \right| |g_{pq} \rangle
    \]
    
    \textbf{• QRAM Update and Pooling:} Update QRAM with $A^{l+1}$ and apply pooling. \\ \hline
    
    \textbf{2. Backward Pass (Quantum Backpropagation):} 
    \vspace{5pt}
    
    \textbf{• Modify the Gradient:} Set to 0 some values of $\frac{\partial L}{\partial Y^{l+1}}$ in QRAM.
    
    \textbf{• Matrix-Matrix Multiplications:} \[
    \left\{
    \begin{aligned}
    &\left(A^l\right)^T \cdot \frac{\partial L}{\partial Y^{l+1}} \cdot \frac{\partial L}{\partial Y^{l+1}} \cdot \left(F^l\right)^T
    \end{aligned}
    \right\}
    \]
    
    \textbf{• Tomography:} Estimate each entry of $\frac{\partial L}{\partial F^l}$ and $\frac{\partial L}{\partial Y^l}$.
    
    \textbf{• Gradient Descent:} \[
    F_{s,q}^{l+1} \leftarrow F_{s,q}^l - \lambda \left( \frac{\partial L}{\partial F_{s,q}^l} \pm \delta \left\| \frac{\partial L}{\partial F^l} \right\|_2 \right)
    \] \\ \hline
    
    \textbf{3. Output:} Updated data matrices $A^{l+1}$ and kernel matrices $F^l$. 

    \includegraphics[width=1\textwidth]{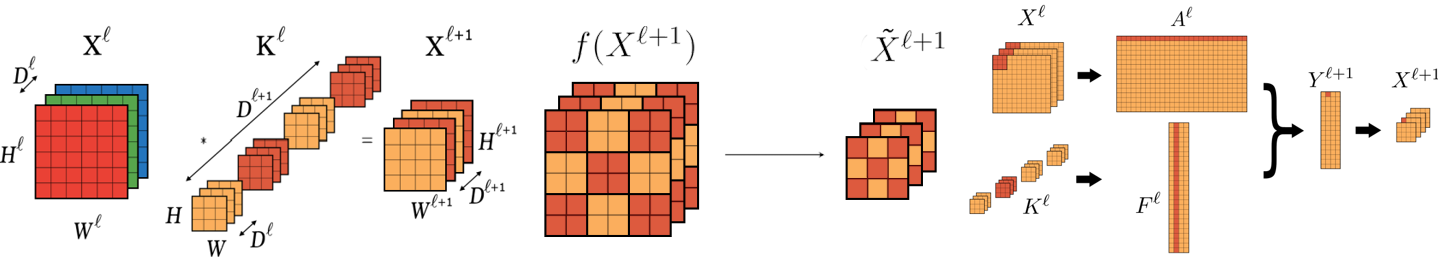} \\ \hline
    \end{tabular}
\end{table}
\vspace{10pt}
As an experiment, the effects of a Quantum Convolutional Neural Network (QCNN) on two samples of monocular earth images are provided in Figure 3, reflecting a detailed testament to adapt classical convolutional neural networks (CNNs) to the quantum setting. This adaptation involves several modifications, including the introduction of quantum sampling after each convolutional layer, the addition of noise to simulate amplitude estimation, the use of a Q-ReLU activation function instead of the traditional ReLU, and the inclusion of noise during backpropagation. The parameters used in this adaptation are critical to the process. The sampling ratio ($\phi$) represents the number of samples drawn during tomography, while the amplitude estimation noise ($\epsilon$) simulates strong noise conditions. The Q-ReLU activation function is defined by a cap value ($K$) that limits the maximum value of pixel intensities, helping to stabilize the training process by controlling the range of activation outputs. The image processing steps are as follows:

\begin{enumerate}[left=0pt]
\item \textbf{Original Image:} The initial step shows the untouched dataset image, serving as a baseline for comparison with subsequent transformations.

\item \textbf{Q-ReLU Activation Function:} The next step applies a Q-ReLU activation with a cap $K$, transforming the image by limiting the maximum pixel intensity values. This transformation stabilizes the training process by preventing extreme activation outputs.

\item \textbf{Noise Introduction:} The third step introduces strong noise to the image, simulating amplitude estimation noise with parameter $\epsilon$. This adds variability and robustness to the model, reflecting real-world noise conditions.

\item \textbf{Quantum Sampling:} The final step performs quantum sampling with a ratio $\phi$, selectively retaining high-intensity pixels based on the sampling threshold. This mimics quantum measurements, preserving only the most significant values and reducing the image to its most informative components.
\end{enumerate}

\begin{figure}[H]
    \centering
    \includegraphics[width=0.49\textwidth]{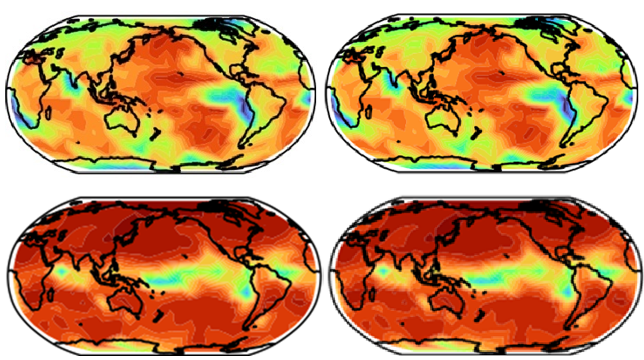}
    \hfill
    \includegraphics[width=0.49\textwidth]{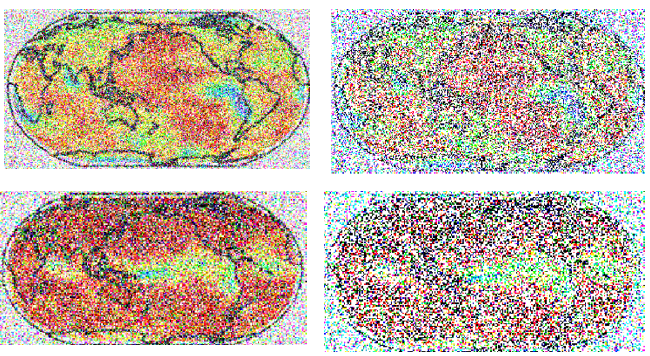}
    \caption{The influences of QCNN on monocular climate images within $K = 10$, $\epsilon = 0.1$, and $\phi = 0.3$.}
    \label{fig:qed}
\end{figure}

As a result of experimenting with the effects of $\phi$, $\epsilon$, and $K$, Figure 4 displays the training performance of Quantum Convolutional Neural Networks (QCNNs) and a classical Convolutional Neural Network (CNN) using various parameters, with the primary metric being Mean Absolute Error (MAE) Loss over epochs. According to the PRECSC figure of merit, with parameters $\epsilon = 0.1$, $K = 10$, and $\phi = 0.1$, the curves represent different QCNN sampling ratios ($\phi = 0.1$ to $0.5$) alongside a CNN. It is observed that the CNN demonstrates smoother and faster convergence compared to the QCNNs. Among the QCNNs, those with higher sampling ratios, such as $\phi = 0.4$, exhibit better performance, reflected in a lower final MAE loss. 

Considering the parameters $\epsilon = 0.1$, $K = 10$, and $\phi = 0.01$, the curves again represent different QCNN sampling ratios ($\phi = 0.1$ to $0.5$) alongside a CNN. Here, the CNN continues to outperform the QCNNs in terms of convergence speed and final MAE loss. Higher sampling ratios in QCNNs tend to yield better performance but still do not surpass the CNN. In the case of parameters $\epsilon = 0.01$, $K = 2$, and $\phi = 0.1$, the CNN outperforms the QCNNs, which show more variability in their performance. On the other hand, turning to parameters of $\epsilon = 0.01$, $K = 2$, and $\phi = 0.4$, it is observed that the QCNN ($\phi = 0.4$) has a distinct advantage, achieving the lowest MAE loss. 

The additional figures with parameters $\epsilon = 0.01$, $K = 10$, $\phi = 0.1$, and $\phi = 0.01$ respectively, continue to show that the QCNN ($\phi = 0.4$) achieves lower MAE loss compared to the CNN across different parameter settings. Overall, the QCNN ($\phi = 0.4$) consistently outperforms the classical CNN across most parameter settings, achieving lower MAE loss and faster convergence. QCNNs show sensitivity to parameter changes, with lower $\epsilon$ and $\phi$ values affecting convergence and final performance. There is noticeable variability in the performance of QCNNs depending on the sampling ratio and other parameters, indicating that fine-tuning is crucial for optimal performance.

\paragraph{2.4.1.1}\textbf{Quantum Dropout Technique}
\vspace{5pt}

In this section, we present a comprehensive discussion of various quantum dropout strategies, as illustrated in Figure 5. Quantum dropout is an essential technique for regularizing quantum neural networks, akin to classical dropout in conventional neural networks. Each strategy employs a unique approach to dropping gates, thereby affecting the network's overall performance and robustness \cite{Scala et al.2023}. In this study, we employ a hybrid policy of given quantum dropout techniques in Figure 5.

\vspace{10pt}

\textbf{Canonical dropout} involves dropping a single rotation gate $R_G$ along with all preceding entangling gates $E_G$ that targeted the particular qubit, and all subsequent entangling gates that used that qubit as a control. As shown in Figure 5, single dropped gates are highlighted by circles/rectangles, with arrows indicating the sequence of dropped gates. This dropout strategy minimizes the network's dependency on specific qubits, potentially enhancing model generalization. The dropping probability $p_G$ is employed together with $p_L$ to obtain the overall dropping probability $p = p_G p_L$.

\vspace{10pt}

\textbf{Canonical-forward dropout} involves dropping a single rotation gate $R_G$ along with all subsequent entangling gates $E_G$ that used that qubit as a control. Illustrated in Figure 5, this method mitigates the influence of future entanglements involving the dropped rotation gate, reducing error propagation through the network. The same dropping probability $p_G$ is utilized.

\vspace{10pt}

\begin{figure}[H]
    \centering
    \includegraphics[width=1\textwidth]{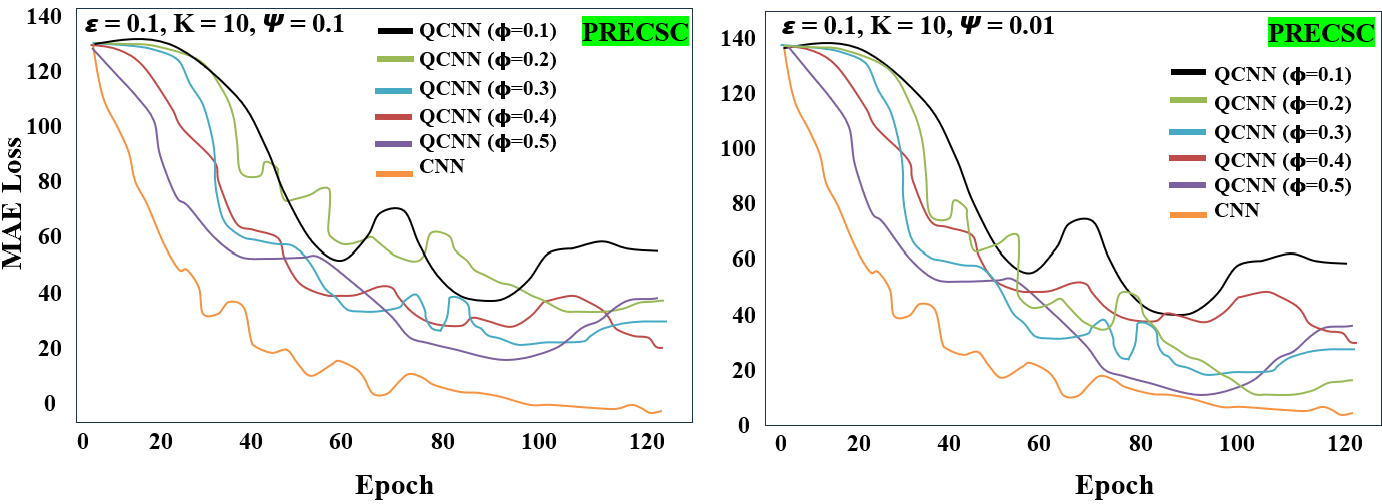}
    \vspace{10pt}
    \centering
    \includegraphics[width=1\textwidth]{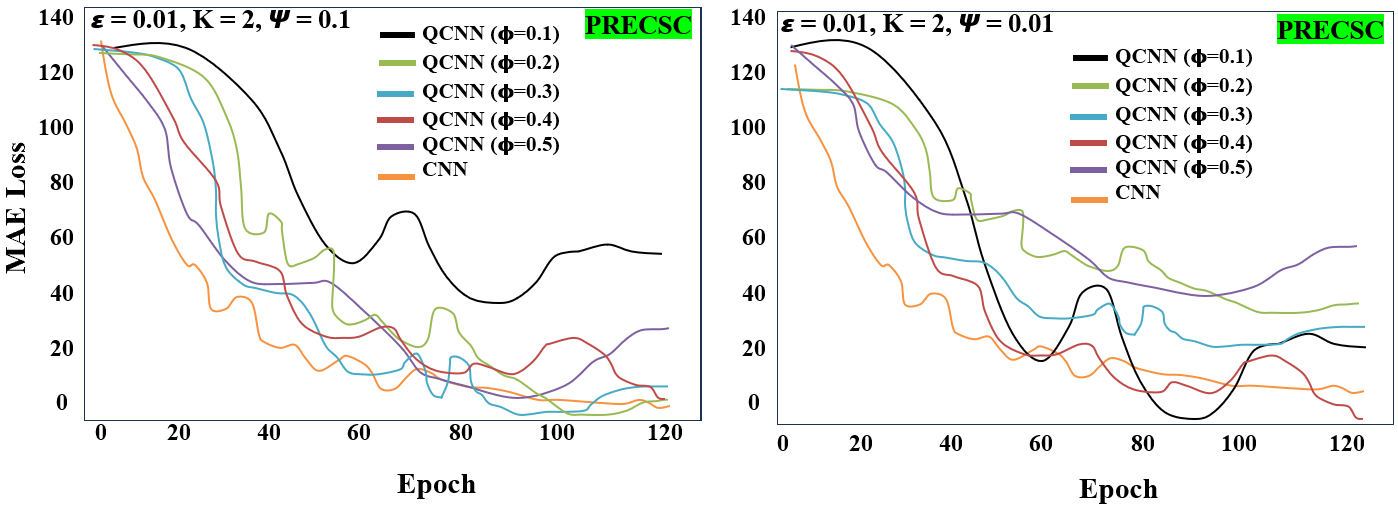}
    \vspace{10pt}
    \centering
    \includegraphics[width=1\textwidth]{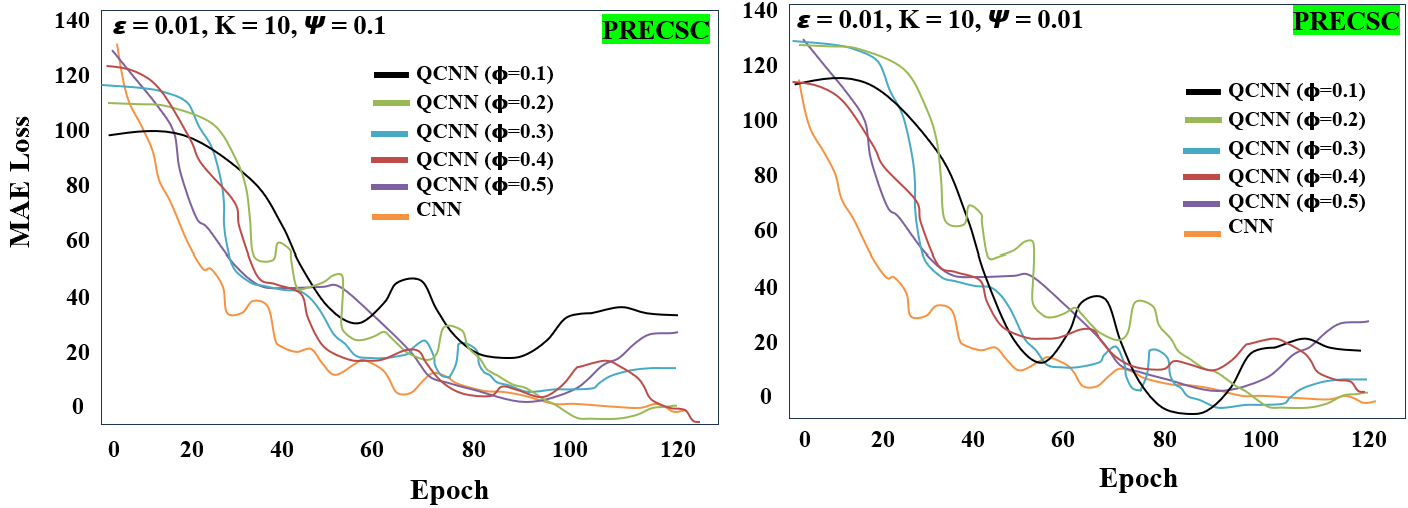}
    \caption{Training results of QCNN model under various conditions for PRECSC output variable.}
    \label{fig:qed}
\end{figure}

\textbf{Independent dropout} works by dropping a single rotation gate $R_G$ and a single entangling gate $E_G$ independently of each other. Figure 5 demonstrates this approach, applying dropout independently to different types of gates, potentially balancing the influence of rotation and entangling gates on the network. This method employs distinct probabilities $p_R$ for rotation gates and $p_E$ for entangling gates.

\vspace{10pt}

\textbf{Rotation dropout} involves dropping single rotation gates $R_G$ alone. As shown in Figure 5, this straightforward approach simplifies the quantum circuit by focusing solely on rotation gates, crucial for qubit state manipulation. The probability $p_R$ is used to determine the dropping of rotation gates.

\vspace{10pt}

\textbf{Entangling dropout} involves dropping single entangling gates $E_G$ alone. Illustrated in Figure 5, this strategy targets entangling gates, pivotal for creating quantum correlations between qubits. Dropping these gates reduces the complexity of the quantum entanglement structure. The probability $p_E$ is used for entangling gate dropout.

\begin{figure}[H]
    \centering
    \includegraphics[width=1\textwidth]{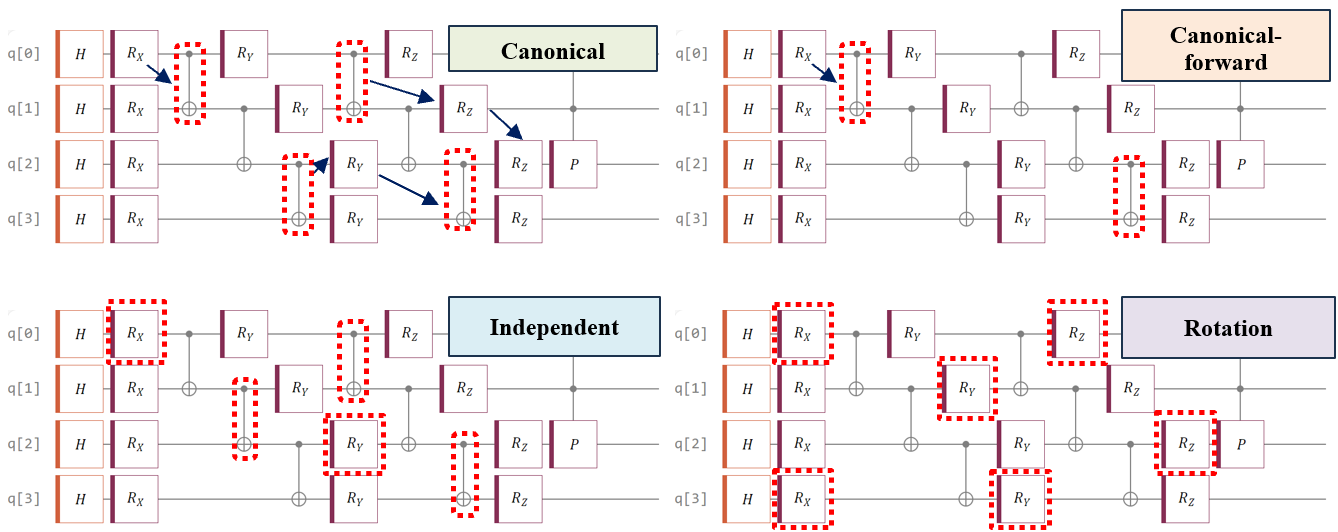}
    \vspace{10pt}
    \centering
    \includegraphics[width=0.5\textwidth]{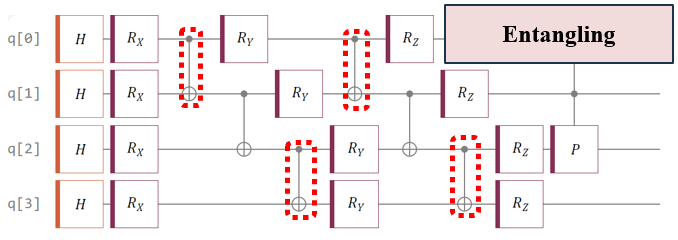}
    \caption{Various quantum dropout techniques utilized in our work.}
    \label{fig:qed}
\end{figure}

\paragraph{2.4.1.2}\textbf{Quantum Activation Function}
\vspace{5pt}

The derivation of quantum-inspired activation functions \cite{Maronese et al.2022}, specifically the Quantum ReLU (Q-ReLU) and the modified Q-ReLU (m-QReLU), using a quantum computing paradigm is elaborated in this section. The process begins with the ReLU and L-ReLU Hilbert state spaces, which represent the state spaces corresponding to the Rectified Linear Unit (ReLU) and Leaky Rectified Linear Unit (L-ReLU) activation functions, respectively. These state spaces undergo entanglement, forming a quantum-based entangled state of solutions. Entanglement combines these states such that the state of one can instantaneously influence the state of the other, reflecting a fundamental property of quantum systems.

From this quantum-based entangled state, the Quantum ReLU (Q-ReLU) is derived. Unlike the classical ReLU, which sets negative inputs to zero, the Q-ReLU assigns these inputs to a quantum state that holds positive solutions. Further, the principle of superposition in quantum mechanics allows for the superposition of states, enabling the function to simultaneously represent positive and negative solutions. This is depicted as the superimposed positive and negative solutions from Q-ReLU. Building on this, the modified Q-ReLU (m-QReLU) is developed by incorporating these superimposed solutions. This modification results in a quantum-inspired activation function with enhanced properties.

Hence, according to Table 2 and Table 3, the respective results, computational time, and accuracy, regarding the proposed Q-ReLU are compared against the most common classical activation functions, and it proves its superiority in terms of accuracy, while the respected computational time significantly increased.

\begin{table}[htbp]
\centering
\caption{Computational time [s] for quantum and classical activation functions.}
\begin{tabular}{lccccccc}
\hline
\textbf{Variable} & \textbf{ReLU} & \textbf{CReLU} & \textbf{Tanh} & \textbf{ELU} & \textbf{SELU} & \textbf{Q-ReLU} \\
\hline
Surface pressure [Pa]    & 1738  & 1718  & 1685  & 1725  & 1676  & \textbf{2755} \\
Insolation [W/m$^2$]     & 1708  & 1700  & 1675  & 1645  & 1678  & \textbf{2658} \\
Latent heat flux [W/m$^2$] & 1644  & 1691  & 1677  & 1695  & 1681  & \textbf{2691} \\
Sensible heat flux [W/m$^2$] & 1711  & 1694  & 1733  & 1705  & 1689  & \textbf{2644} \\
Surface temperature [K]  & 1701  & 1724  & 1655  & 1715  & 1702  & \textbf{2745} \\
Precipitation [mm/day]   & 1688  & 1692  & 1690  & 1658  & 1722  & \textbf{2683} \\
Surface wind speed [m/s] & 1694  & 1666  & 1653  & 1636  & 1711  & \textbf{2777} \\
Longwave radiation [W/m$^2$] & 1677  & 1689  & 1721  & 1686  & 1700  & \textbf{2633} \\
\hline
\end{tabular}
\end{table}

\begin{table}[htbp]
\centering
\caption{Accuracy [0-1] for quantum and classical activation functions.}
\begin{tabular}{lccccccc}
\hline
\textbf{Variable} & \textbf{ReLU} & \textbf{CReLU} & \textbf{Tanh} & \textbf{Q-ReLU} & \textbf{SELU} & \textbf{ELU} \\
\hline
Surface pressure [Pa]    & 0.944  & 0.952  & 0.933  & \textbf{0.995}  & 0.920  & 0.916 \\
Insolation [W/m$^2$]     & 0.829  & 0.909  & 0.914  & \textbf{0.984}  & 0.822  & 0.875 \\
Latent heat flux [W/m$^2$] & 0.912  & 0.922  & 0.913  & \textbf{0.998}  & 0.920  & 0.914 \\
Sensible heat flux [W/m$^2$] & 0.924  & 0.931  & 0.922  & \textbf{0.994}  & 0.919  & 0.936 \\
Surface temperature [K]  & 0.933  & 0.927  & 0.927  & \textbf{0.993}  & 0.920  & 0.911 \\
Precipitation [mm/day]   & 0.937  & 0.928  & 0.938  & \textbf{0.989}  & 0.925  & 0.932 \\
Surface wind speed [m/s] & 0.917  & 0.934  & 0.934  & \textbf{0.996}  & 0.961  & 0.957 \\
Longwave radiation [W/m$^2$] & 0.833  & 0.856  & 0.866  & \textbf{0.949}  & 0.852  & 0.883 \\
\hline
\end{tabular}
\end{table}

\subsubsection{Quantum Multilayer Perceptron (QMLP)}
According to depicted architecture in Figure 7, it represents a QMLP developed for advanced data processing tasks. This model integrates quantum computing principles with classical data processing techniques to achieve enhanced computational capabilities. At the outset, the input section comprises a series of earth images that represent diverse data distributions, potentially including geographical and spatial-temporal datasets. These input images serve as the initial data for subsequent quantum encoding. In the quantum encoding phase, classical data is transformed into quantum states suitable for processing by quantum circuits. This step employs angle and amplitude encoding methods to convert each pixel value or data point from the images into qubits. Specifically, classical encoding (C) handles values in the range of 0 to 1, while quantum encoding (Q) represents quantum states such as $\lvert 0 \rangle$ and $\lvert 1 \rangle$. The encoding translates the classical information into a quantum format, laying the groundwork for further quantum processing. Herein, four encoded qubits with different data distributions ranging between 0 and 1 are visualized in Figure 7.
 
Following the encoding, the quantum neural networks (QNNs) process the encoded quantum states. The QNNs consist of multiple quantum nodes (Q1, Q2, Q3, Q4), each performing specific quantum computations and transformations on the qubits. These nodes represent layers in the quantum neural network, applying a series of quantum gates that manipulate the quantum states to extract meaningful patterns and relationships within the data. The quantum nodes are interconnected in a manner analogous to a classical multilayer perceptron (MLP) but operating entirely within the quantum domain. This Quantum MLP leverages quantum entanglement and other quantum operations, facilitating complex transformations and interactions among the quantum states. The detailed QMLP ablation study is also provided as below.

As shown in Figure 7, it illustrates a comprehensive Quantum Multilayer Perceptron (QMLP) architecture specifically designed for analyzing Earth and climate data. This model is composed of several critical components and processes that work together to leverage quantum computing capabilities for enhanced data processing and analysis.
The input data consists of various Earth and climate datasets, visually represented by three sample images on the left side of the diagram. These images encapsulate information about multiple climatic parameters across different geographical regions. To process this data using quantum computing, it first undergoes a transformation through Quantum Encoding techniques. The two primary encoding methods used are Angle Encoding and Amplitude Encoding. Angle Encoding maps data values into the angles of quantum states, whereas Amplitude Encoding maps data values into the amplitudes of these states. This transformation utilizes a combination of classical (C) and quantum (Q) components, effectively converting binary values (0 and 1) into corresponding qubit states $\lvert 0 \rangle$ and $\lvert 1 \rangle$. This step is crucial as it prepares the classical data for quantum processing \cite{Friedrich et al.2024,Shao2018}.

Once the data is encoded into quantum states, it is fed into a series of Quantum Neural Networks (QNNs). These networks consist of qubits, labeled Q1, Q2, Q3, and Q4 in the diagram. Each qubit forms part of a quantum circuit, performing a sequence of quantum operations such as quantum gates, which manipulate the quantum states based on the input data. The interaction between qubits through these gates allows the QNNs to leverage quantum phenomena like superposition and entanglement, enabling complex data transformations that are not possible with classical neural networks. The processed quantum states from the QNNs are then passed into the Quantum Multilayer Perceptron (QMLP). The QMLP consists of multiple layers of quantum neurons, which are interconnected through weights and biases in a manner similar to classical multilayer perceptron but within the quantum domain. Each layer of the QMLP performs quantum operations on the input states, transforming them as they propagate through the network. The key advantage of QMLPs over classical MLPs lies in their ability to utilize quantum parallelism and entanglement, potentially offering more powerful and efficient computations. More schematically, a visual representation of a quantum circuit associated with the QMLP architecture has been provided in Figure 6.

\begin{figure}[H]
    \centering
    \includegraphics[width=0.8\textwidth]{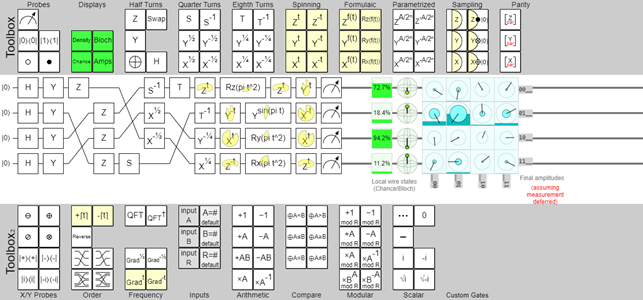}
    \caption{The detailed schematic of the quantum circuit utilized in QMLP architecture.}
    \label{fig:qed}
\end{figure}

The final step involves measuring the output of the Quantum MLP. Quantum measurement devices are used to convert the quantum states back into classical information, which can be interpreted and analyzed. The output includes various climatic parameters and their corresponding values, such as Net Shortwave Radiation (NETSW), Downward Longwave Radiation (FLWDS), Convective Snowfall Rate Water Equivalent (PRECSC), Convective Precipitation Rate (PRECC), Surface Solar Radiation (SOLS), Low-Level Solar Radiation (SOLL), Downward Solar Radiation (SOLSD), and Downward Longwave Radiation (SOLLD). These outputs are critical for understanding and predicting climatic changes and behaviors, providing valuable insights for climate science.

The bottom section of Figure 7 visualizes the quantum states using Bloch spheres. Each Bloch sphere represents the state of a qubit, illustrating the probabilities of the qubit being in state $\lvert 0 \rangle$ or $\lvert 1 \rangle$, as well as their superpositions. The Bloch spheres offer an intuitive way to understand the state transformations that occur during the processing in QNNs and QMLP. The visualization shows how input data is mapped onto the quantum states, highlighting the quantum encoding and processing steps. The general state of a qubit $\lvert \psi \rangle$ can be described by the equation:
\[
\lvert \psi \rangle = \cos(\theta / 2) \lvert 0 \rangle + e^{i \phi} \sin(\theta / 2) \lvert 1 \rangle
\]
where $\theta$ and $\phi$ are the spherical coordinates on the Bloch sphere. The z-axis represents the computational basis states $\lvert 0 \rangle$ and $\lvert 1 \rangle$, while the x- and y-axes represent the superpositions of these states, such as $\lvert + \rangle = (\lvert 0 \rangle + \lvert 1 \rangle) / 2$ and $\lvert - \rangle = (\lvert 0 \rangle - \lvert 1 \rangle) / 2$. Pure states are depicted as points on the surface of the sphere, whereas mixed states, which are probabilistic combinations of pure states, would lie inside the sphere (though these are not depicted in this Figure).

Each Bloch sphere at the bottom of Figure 7 represents the quantum states of the qubits after encoding and processing through the Quantum Neural Networks (QNNs) and the Quantum MLP. Initially, qubits are in a definite state, usually $\lvert 0 \rangle$, depicted at the north pole of the Bloch sphere. As data is encoded into the quantum states, the qubits move from their initial state to various points on the Bloch sphere, depending on the encoding scheme (angle or amplitude encoding). The encoded state reflects the information from the input data. For instance, if amplitude encoding is used, the probability amplitudes of the qubit states correspond to the input data values.
\vspace{10pt}

\begin{figure}[H]
    \centering
    \includegraphics[width=1\textwidth]{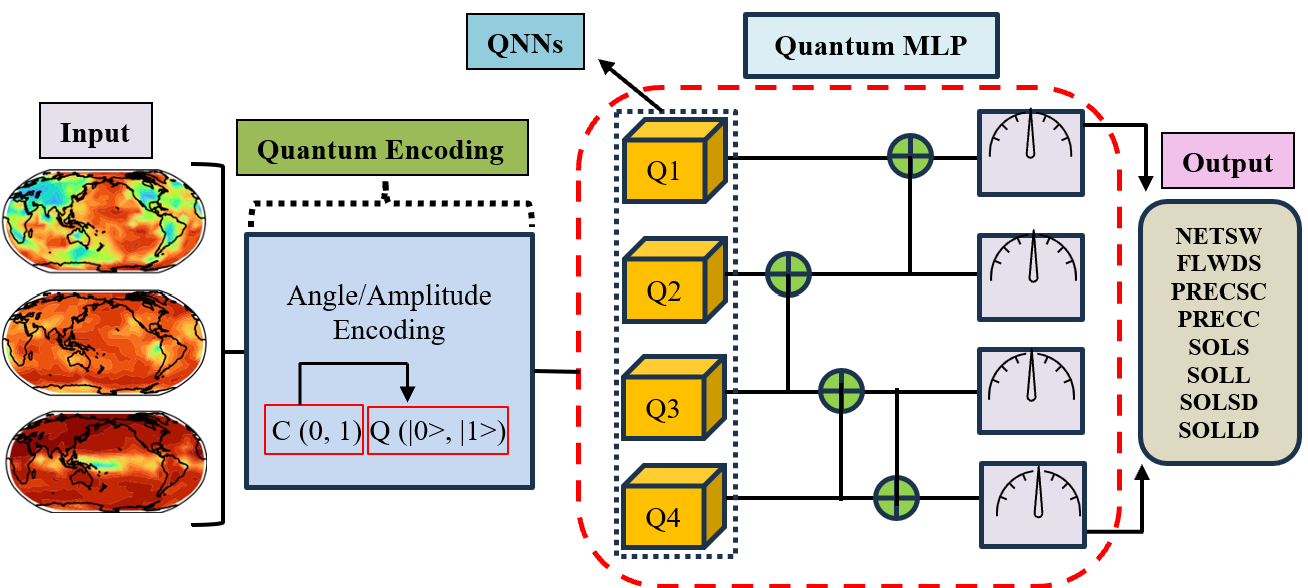}
    \vspace{10pt}
    \centering
    \includegraphics[width=1\textwidth]{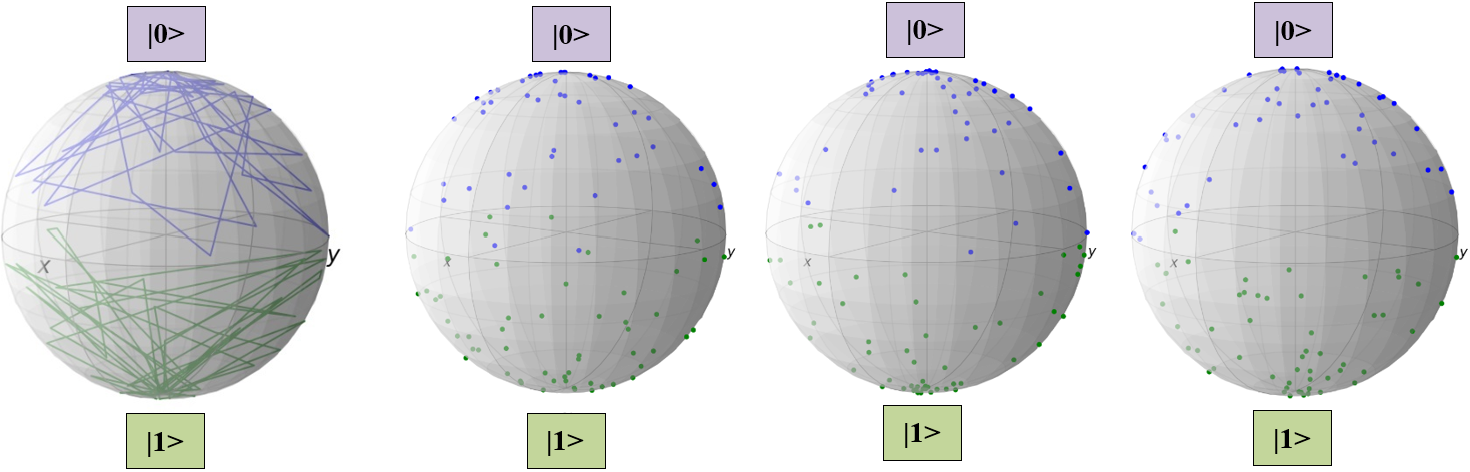}
    \caption{The detailed schematic of the QMLP model along with its quantum components and encoded qubits.}
    \label{fig:qed}
\end{figure}

As a part of the ablation study in Figure 8, the performance of Quantum Machine Learning Perceptron (QMLP) models was evaluated under various configurations of qubits (Q) and layers (L). The results indicate that using multiple shallow quantum circuits (ensemble learning) consistently outperforms or matches the performance of deeper single-circuit models, effectively addressing the challenges of vanishing gradient (VG) and cost function concentration (CFC).

The analysis shows that m-QReLU generally provides lower loss and improved stability compared to Q-ReLU, particularly in configurations with higher qubit counts and layers. For instance, with n-Q = 6 and n-L = 3, m-QReLU achieves a significantly lower loss, suggesting better efficiency and robustness. This trend continues across various configurations, with m-QReLU maintaining a lower loss and demonstrating resilience to parameter initialization, especially as the model complexity increases.

Notably, as the depth and number of qubits increase, both activation functions tend to perform similarly, but m-QReLU still holds a slight edge in most cases. This indicates that ensemble learning enhances model capacity and mitigates VG and CFC issues, making it a more effective approach for optimizing QMLP models.
\vspace{10pt}

\begin{figure}[H]
    \centering
    \includegraphics[width=1\textwidth]{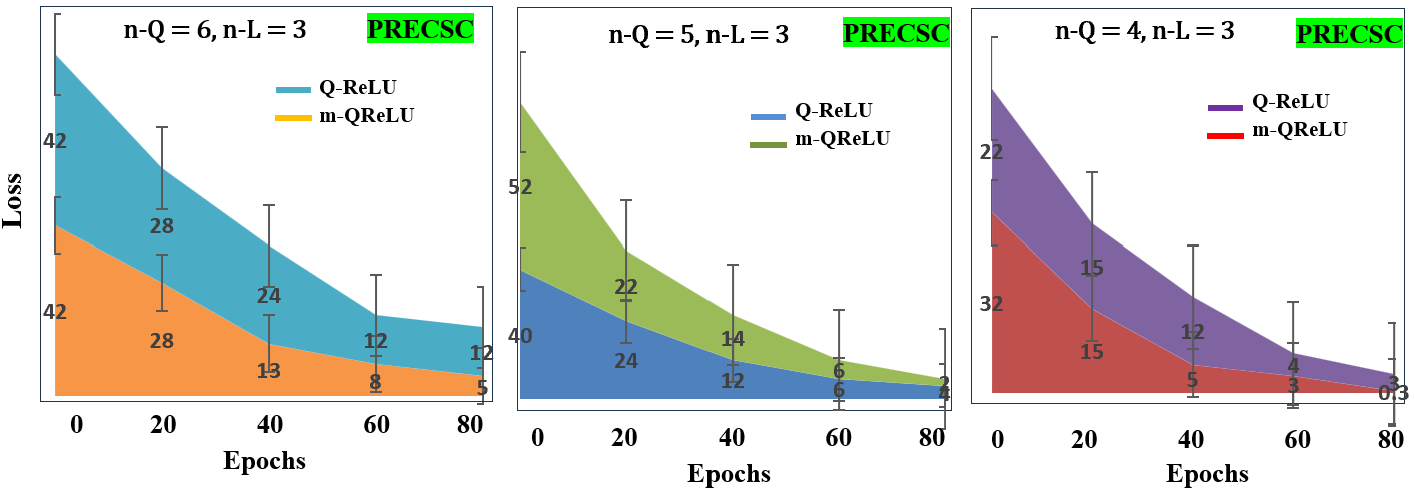}
    \vspace{10pt}
    \centering
    \includegraphics[width=1\textwidth]{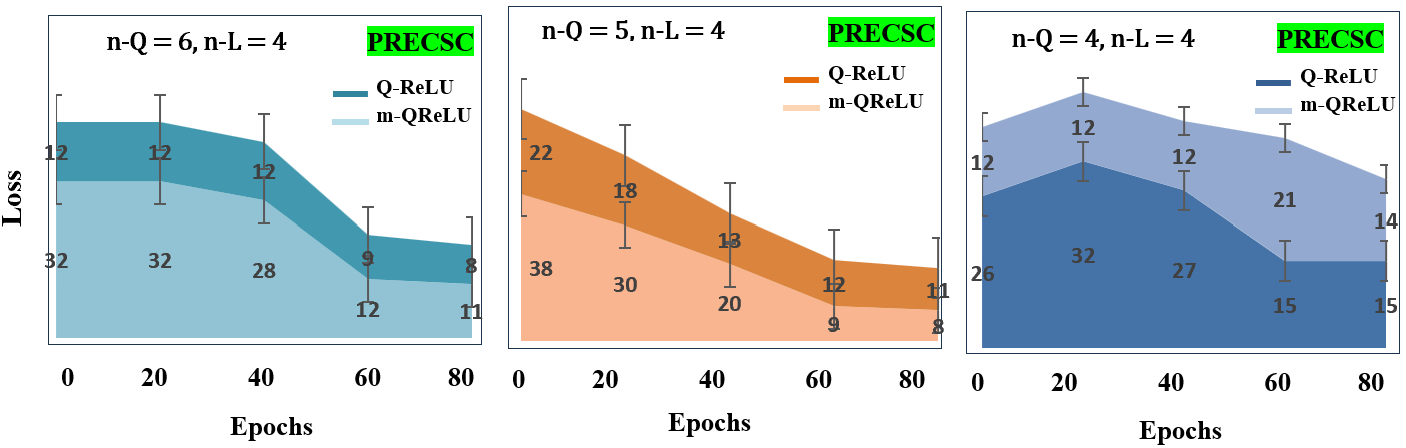}
    \vspace{10pt}
    \centering
    \includegraphics[width=1\textwidth]{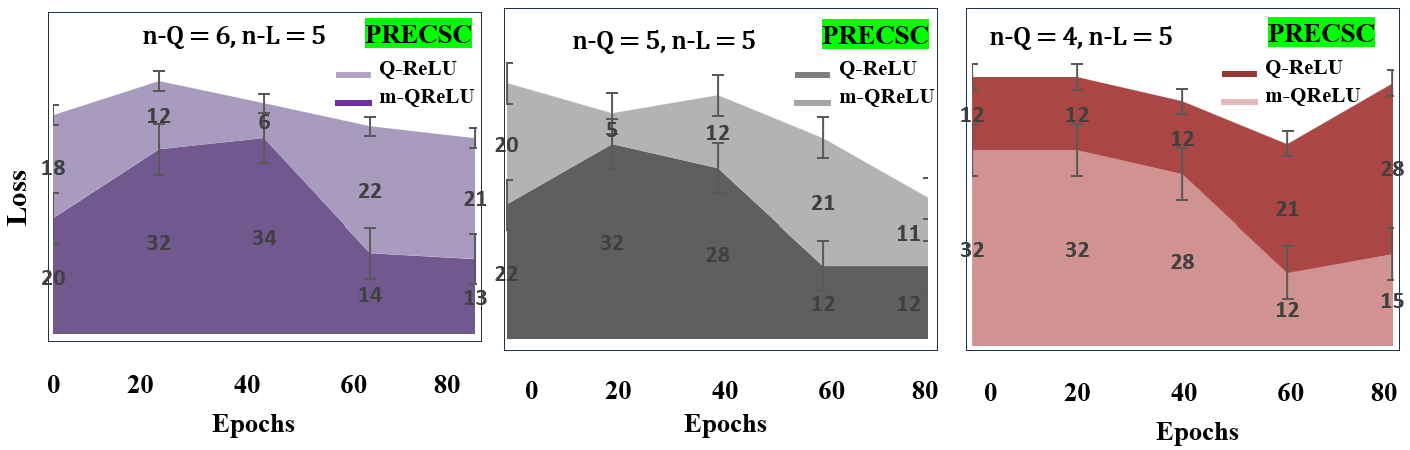}
    \caption{Ablation study of quantum components associated with the QMLP framework.}
    \label{fig:qed}
\end{figure}

\subsubsection{Quantum Encoder-Decoder (QED)}
The QED architecture illustrated in Figure 9 is devoted to advanced quantum counterpart of classical autoencoder (CAE) tasks, incorporating quantum computing elements to enhance processing capabilities. The inputs are processed by the encoder, denoted as $G_\theta$, which compresses the input information into a lower-dimensional latent vector ($Z$). The encoder generates a probability distribution $P(Z|X)$, indicating the likelihood of the latent vector given the input data. Following the encoding process, the latent vector ($Z$) is passed to the decoder, denoted as $F_\theta$. The decoder's task is to reconstruct the input data from the latent representation, generating an output $\hat{X}$ from $Z$. This process produces a probability distribution $P(\hat{X}|Z)$, representing the likelihood of the reconstructed data given the latent vector. More specifically, a detailed view of the quantum circuit used in both the encoder and decoder is displayed with the preparation of four input qubits ($Q1, Q2, Q3, Q4$) to represent the input data. These qubits are initialized into a specific quantum state $\lvert \Psi_{in}^i \rangle$ through an initialization operation $S_i$. During the encoding process, some qubits are frozen, preserving their states to maintain certain information intact while other parts of the qubits undergo transformation. A parameterized unitary transformation $U_\epsilon (\theta)$ is then applied to the active qubits, encoding the input data into the latent space qubits $\lvert \psi_{in}^i \rangle$. These latent space qubits hold the encoded information, which is subsequently transformed back into a form suitable for decoding by another unitary transformation $U_D (\theta)$. The output qubits $\lvert \psi_{out}^i \rangle$ are measured and processed to yield the final output states $S_i^{+}$.
The performance of this quantum encoder-decoder model is guided by a loss function $L(\theta)$, defined as \[\frac{1}{M} \sum_{j}^{M} \lvert 1 - S_i^{+} \rvert,\] where $M$ represents the number of measurements. This loss function measures the discrepancy between the expected and actual outputs, providing a basis for optimizing the model parameters $\theta$ to minimize reconstruction errors. More technical details regarding the QED model have been elaborated as following.

The diagram in Figure 9 represents an advanced Quantum Encoder-Decoder system designed for efficiently compressing and reconstructing environmental and climatic data using quantum computing principles. Below is a comprehensive analysis of each component and their interactions within the system.

\paragraph{2.4.3.1}\textbf{Input Data}
\vspace{5pt}

The input to the system comprises multispectral images or other relevant datasets that capture various environmental parameters. These could include temperature maps, precipitation levels, humidity distributions, and other geospatial data vital for comprehensive climate analysis.
\begin{enumerate}[left=0pt]
    \item \textbf{Data Representation:} The input data is typically represented as high-dimensional arrays (e.g., images with multiple channels corresponding to different spectral bands).
    \item \textbf{Normalization:} Prior to feeding into the quantum encoder, the data is likely normalized to ensure consistency and enhance the quantum operations' efficiency.
\end{enumerate}

\paragraph{2.4.3.2}\textbf{Quantum Encoder ($G_{\theta}$)}
\vspace{5pt}

The encoder \cite{Wu et al.2024,Rao et al.2023} component is tasked with compressing the input data into a more manageable, lower-dimensional representation. This process includes several crucial steps:
\begin{enumerate}[left=0pt]
    \item \textbf{Initial Encoding:} The high-dimensional input data $X$ is encoded into quantum states. This step involves mapping the classical data onto quantum bits (qubits), enabling quantum processing.
    \item \textbf{Quantum Operations:} Various quantum gates and operations are applied to transform the input data into a latent representation $Z$. These operations exploit quantum parallelism and entanglement to capture complex data features.
    \item \textbf{Latent Space Representation:} The resultant latent vector $Z$ has a specified latent diameter of 5, meaning the data is compressed into a 5-dimensional space, balancing dimensionality reduction and information retention.
\end{enumerate}

\paragraph{2.4.3.3}\textbf{Latent Space}
\vspace{5pt}

The latent space is a crucial intermediate representation where the high-dimensional input data is condensed into a lower-dimensional form. Key aspects include:
\begin{enumerate}[left=0pt]
    \item \textbf{Dimensionality Reduction:} Reducing the data's dimensions helps in simplifying the subsequent processing steps while retaining the essential features.
    \item \textbf{Feature Extraction:} The encoder extracts significant features from the input data, encapsulating them in the latent vector $Z$.
\end{enumerate}

\paragraph{2.4.3.4}\textbf{Quantum Decoder ($F_{\theta}$)}
\vspace{5pt}

The decoder's \cite{Rea et al.2024} role is to reconstruct the original input data from the compressed latent vector $Z$. This involves:
\begin{enumerate}[left=0pt]
    \item \textbf{Data Reconstruction:} Using quantum operations, the decoder transforms the latent vector back into a higher-dimensional form that approximates the original input data $X$.
    \item \textbf{Output Generation:} The decoder produces various environmental indicators such as NETSW, FLWDS, PRECSC, PRECC, SOLS, SOLL, SOLSD, SOLLD. These indicators are derived from the reconstructed data, providing valuable insights into climatic and environmental conditions.
\end{enumerate}

\paragraph{2.4.3.5}\textbf{Quantum Circuit Details}
\vspace{5pt}

The quantum circuit is a pivotal component, leveraging quantum mechanics to perform the encoding and decoding processes. It includes:
\begin{enumerate}[left=0pt]
    \item \textbf{Input Qubits:} The input data is represented by initial qubits (Q1, Q2, Q3, Q4), encoding the classical data into quantum states.
    \item \textbf{State Preparation ($S_i$):} This unit prepares the quantum states ($|\psi^i_{in}\rangle$) by encoding the input data into qubits.
    \item \textbf{Freezing and Active Qubits:}
    \begin{itemize}[left=0pt]
    \item \textbf{Freezing Qubits:} Some qubits remain in a fixed state during certain operations to preserve specific information.
    \item \textbf{Active Qubits:} Other qubits undergo various quantum transformations, actively participating in encoding and decoding processes.
    \end{itemize}
    \item \textbf{Unitary Transformations ($U_{\epsilon}(\theta)$):} 
    \begin{itemize}[left=0pt]
        \item \textbf{Encoding Transformation:} This unitary transformation processes the latent space qubits, encoding the input qubit states to correspond to the input data values.
    \end{itemize}
    \item \textbf{Decoder Unitary ($U_D(\theta)$):} 
    \begin{itemize}[left=0pt]
        \item \textbf{Decoding Transformation:} This transformation decodes the latent space qubits, transforming them back into a representation suitable for output qubits.
    \end{itemize}
    \item \textbf{Output Qubits:} The quantum states ($|\psi^i_{out}\rangle$) are measured, converting the quantum data back into classical data. The output qubits yield the reconstructed data. 
\end{enumerate}

\paragraph{2.4.3.6}\textbf{Loss Function}
\vspace{5pt}

The system employs a loss function $L(\theta) = \frac{1}{M} \sum_{j}^{M} |1 - S^+_i|$, where $M$ is the number of samples. This function measures the difference between the original and reconstructed data. Key points include:
\begin{enumerate}[left=0pt]
    \item \textbf{Error Measurement:} The loss function quantifies the reconstruction error, guiding the optimization process.
    \item \textbf{Parameter Optimization:} By minimizing the loss, the parameters $\theta$ are adjusted to improve the accuracy of the reconstruction.
    \item \textbf{Training Process:} The optimization involves iterative adjustments to the quantum operations to minimize the reconstruction error.
\end{enumerate}

To further proceed with the ablation study of the QED model, as can be observed in Figure 10, the main focus is on the impact of the regularization parameter $\beta$ on the fidelity and regularization loss metrics across different datasets and model configurations. The primary objective is to understand the trade-offs between reconstruction quality and latent space regularization and to evaluate the role of auxiliary qubits in this context.

\paragraph{2.4.3.7}\textbf{Fidelity Analysis}
\vspace{5pt}

Fidelity is a measure of the reconstruction quality of the model. Across all datasets, an increase in $\beta$ generally results in a decrease in fidelity, indicating that higher regularization parameters negatively impact the model's ability to reconstruct the original data accurately. For example, in the NETSW dataset, fidelity without auxiliary qubits drops from approximately 0.95 at $\beta=0$ to about 0.75 at $\beta=6$. A similar trend is observed in the SOLL dataset, where fidelity decreases from around 0.93 to 0.62 over the same range of $\beta$. Models without auxiliary qubits tend to maintain higher fidelity compared to those with auxiliary qubits. This suggests that while auxiliary qubits enhance regularization, they do so at the expense of reconstruction quality.

\begin{figure}[H]
    \centering
    \includegraphics[width=1\textwidth]{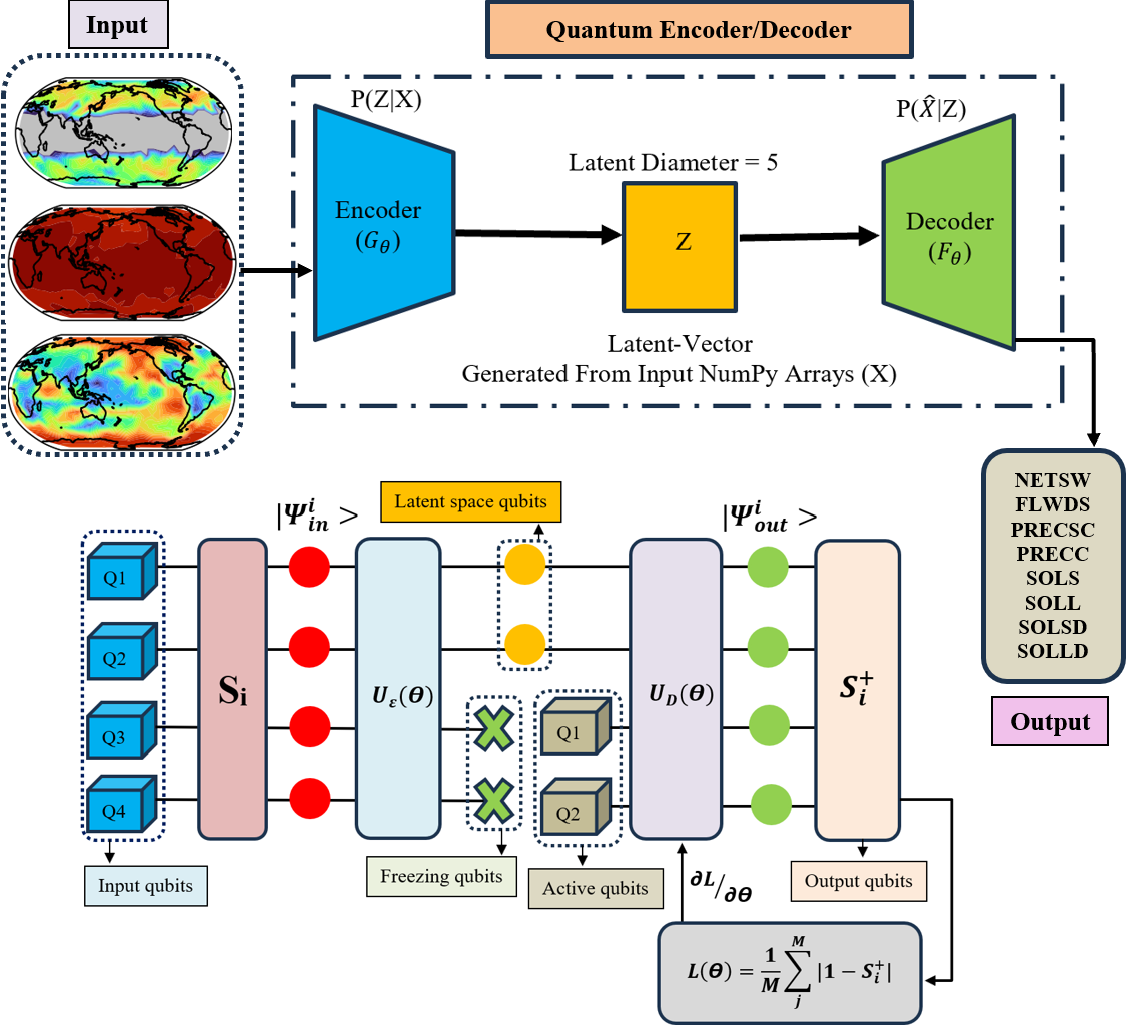}
    \caption{The detailed schematic of the QED model along with its quantum components.}
    \label{fig:qed}
\end{figure}

\paragraph{2.4.3.8}\textbf{Regularization Loss Analysis}
\vspace{5pt}

Regularization loss is a measure of how well the model manages the latent space. As $\beta$ increases, regularization loss also increases, indicating stronger regularization effects. For instance, in the FLWDS dataset, the regularization loss without auxiliary qubits increases from about 0.2 to 0.9 as $\beta$ goes from 0 to 6, while with auxiliary qubits, it increases from 0.15 to 0.85. Models with auxiliary qubits consistently show lower regularization loss compared to those without, highlighting the effectiveness of auxiliary qubits in better managing the latent space. The presence of auxiliary qubits allows the model to decouple the reconstruction and regularization processes, leading to improved regularization without excessively compromising fidelity.

\paragraph{2.4.3.9}\textbf{Trade-offs and Interactions}
\vspace{5pt}

The interaction between fidelity and regularization loss reveals important trade-offs. Higher $\beta$ values enhance regularization but reduce fidelity. For example, in the PRECSC dataset, fidelity drops from around 0.9 to 0.7, while the regularization loss increases from 0.3 to 0.8 as $\beta$ increases. This trade-off necessitates careful tuning of $\beta$ to balance reconstruction quality and regularization according to the specific requirements of the application. The inclusion of auxiliary qubits introduces an additional layer of complexity, offering improved regularization capabilities at the cost of lower fidelity. For instance, in the SOLSD dataset, with auxiliary qubits, fidelity remains higher at lower $\beta$ values but shows a more pronounced decrease as $\beta$ increases compared to the setup without auxiliary qubits. This suggests that auxiliary qubits can be particularly beneficial when the primary goal is to achieve better latent space management, even if it means accepting a slight decrease in reconstruction accuracy.

\paragraph{2.4.3.10}\textbf{Dataset-Specific Observations}
\vspace{5pt}

While the overall trends of decreasing fidelity and increasing regularization loss with higher $\beta$ values hold across all datasets, the specific impact varies. In the PRECC dataset, for instance, the model with auxiliary qubits maintains a higher fidelity at lower $\beta$ values (around 0.85 at $\beta = 0.5$) but decreases significantly with higher $\beta$ values (down to 0.6 at $\beta = 6$). This variation underscores the importance of dataset-specific tuning of $\beta$ and the consideration of auxiliary qubits. Certain datasets may benefit more from the enhanced regularization provided by auxiliary qubits, while others may prioritize higher fidelity.

\paragraph{2.4.3.11}\textbf{Implications for Model Design}
\vspace{5pt}

The analytical results emphasize the importance of balancing $\beta$ and the use of auxiliary qubits in designing quantum autoencoder (QAE) models. For applications requiring high reconstruction quality, lower $\beta$ values and the exclusion of auxiliary qubits might be preferable. Conversely, for applications where latent space regularization is critical, higher $\beta$ values and the inclusion of auxiliary qubits can offer significant advantages. To continue with the ablation analysis of the results shown in Tables 4-6, we need to evaluate the performance of different loss functions (Fidelity, Wasserstein, and Jensen-Shannon Divergence) across three regularization strengths ($\beta = 0$, $\beta = 3$, and $\beta = 6$) for various algorithms (NETSW, FLWDS, PRECSC, PRECC, SOLS, SOLL, SOLSD, SOLLD). For $\beta = 0$, the regularization term is not applied, focusing purely on reconstruction fidelity. The NETSW algorithm shows high Fidelity and Wasserstein performance (approximately 0.938 and 0.945), indicating strong reconstruction capabilities without regularization. Its JSD is also high, signifying low divergence. The FLWDS algorithm has the highest Fidelity (around 0.991) but a lower Wasserstein (around 0.902), suggesting excellent reconstruction in terms of fidelity but a slight drop when considering Wasserstein distance, potentially due to the method's nature. PRECSC also shows high Fidelity and Wasserstein values, indicating good reconstruction performance. PRECC, however, is lower than PRECSC in both metrics, suggesting lesser performance. The SOLS algorithm demonstrates high performance across all metrics, especially JSD, and SOLL and SOLLD both show high performance with minor variances.

\begin{figure}[H]
    \centering
    \includegraphics[width=1\textwidth]{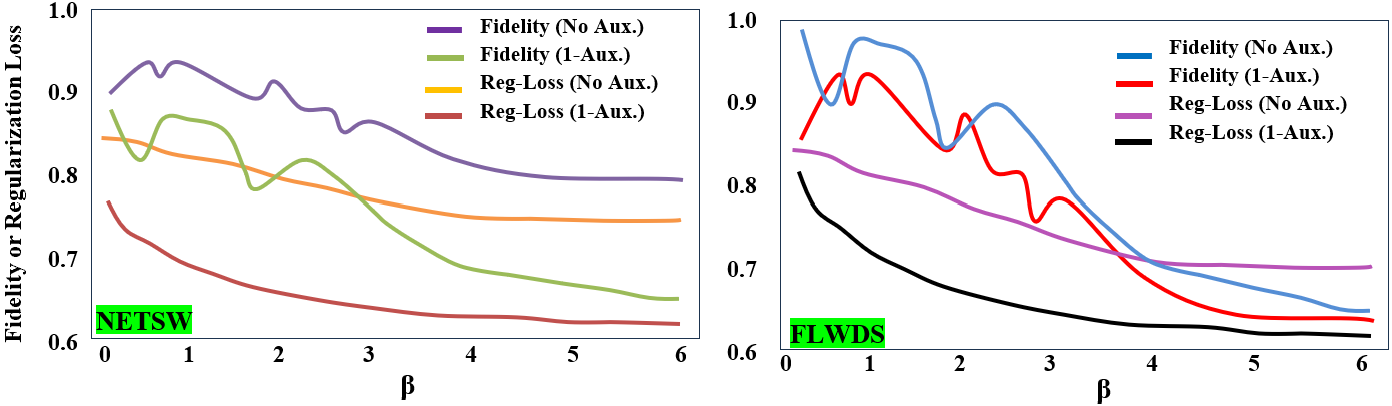}
    \centering
    \includegraphics[width=1\textwidth]{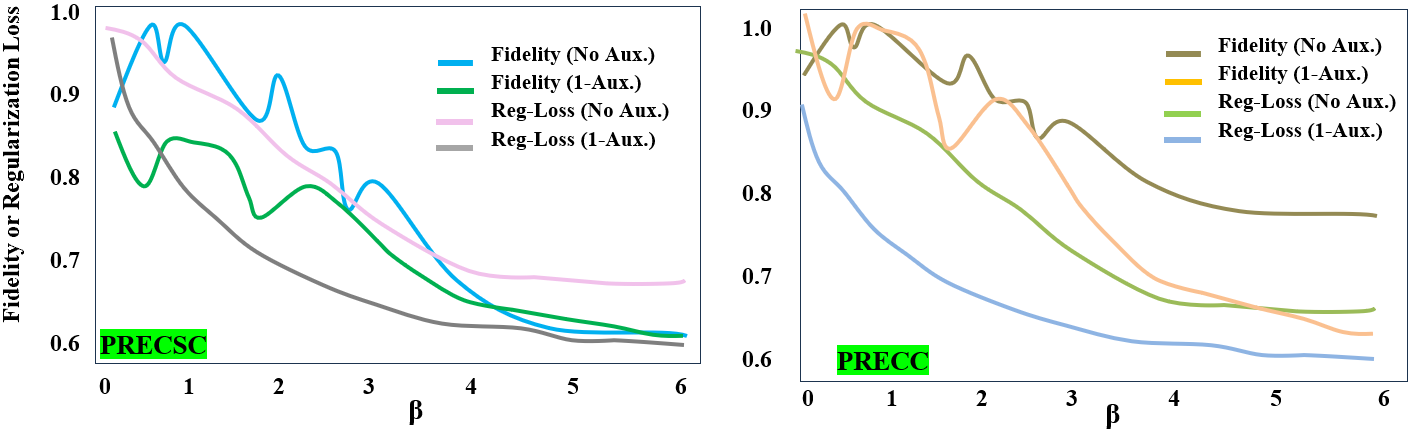}
    \centering
    \includegraphics[width=1\textwidth]{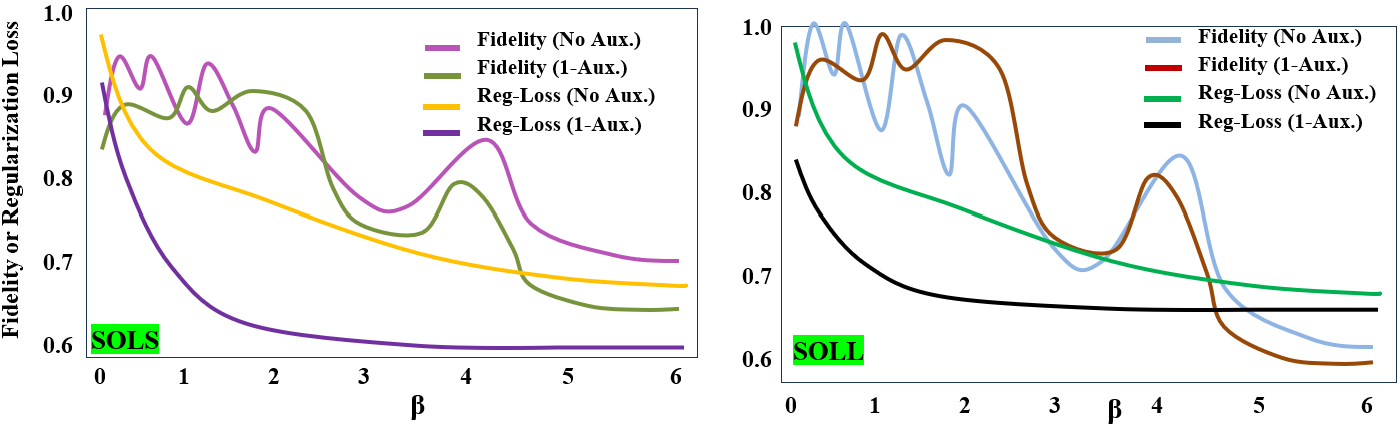}
    \centering
    \includegraphics[width=1\textwidth]{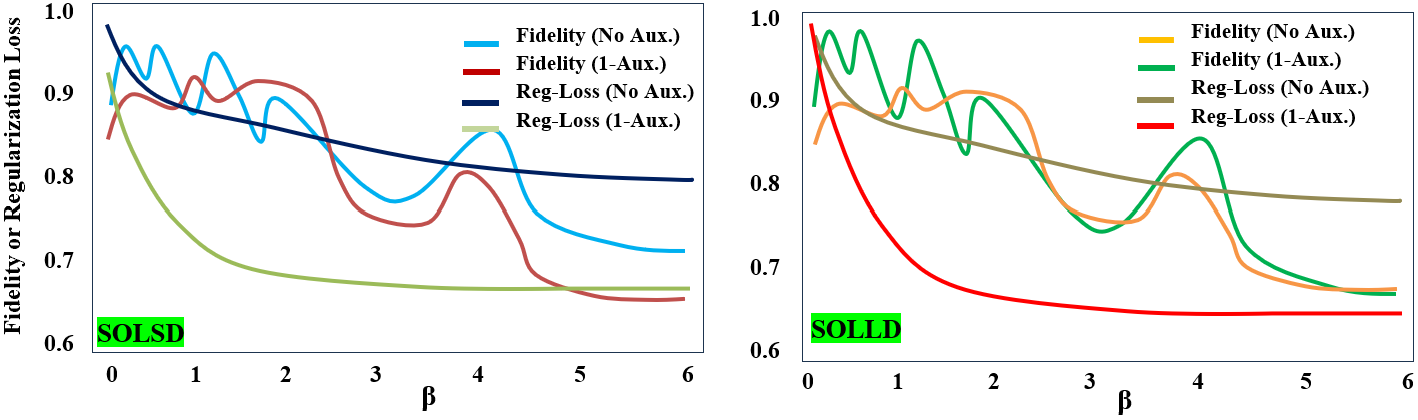}
    \caption{The ablation study of auxiliary qubits for QED model.}
    \label{fig:qed}
\end{figure}

Introducing regularization with $\beta = 3$ begins to balance reconstruction with generalization. The NETSW algorithm shows a noticeable drop in all metrics, especially Fidelity (around 0.743), indicating that this method struggles with maintaining performance when regularization is introduced. FLWDS experiences a drop in Fidelity (around 0.805) but performs better than NETSW and shows improved performance in JSD and Wasserstein compared to NETSW. PRECSC maintains relatively high values (around 0.896 Fidelity), indicating robustness against regularization. PRECC follows a similar pattern to NETSW, struggling with regularization. SOLS shows a moderate drop but maintains decent values, suggesting robustness, and SOLL and SOLLD show better resilience compared to SOLS.
\vspace{10pt}

Higher regularization with $\beta = 6$ emphasizes generalization over pure reconstruction. The NETSW algorithm sees a further drop in performance, particularly in Fidelity (around 0.632), indicating significant struggle under high regularization. FLWDS shows consistent but reduced performance, still better than NETSW. PRECSC once again shows resilience with better scores compared to NETSW and FLWDS. PRECC's performance degrades further, aligning with previous trends. SOLS continues to show moderate performance, while SOLL and SOLLD demonstrate resilience similar to their performance in lower regularization settings. In summary, the analysis reveals several key findings. Algorithms like PRECSC and SOLLD demonstrate robustness to regularization, maintaining relatively high Fidelity, Wasserstein, and JSD scores across different $\beta$ values. In contrast, NETSW and PRECC show significant drops in performance as $\beta$ increases, indicating less robustness to regularization. Fidelity typically shows the highest scores but also the most significant drops with increasing $\beta$. Wasserstein and JSD provide a more balanced view, with JSD often showing higher resilience.
\vspace{30pt}

\begin{table}[htbp]
\centering
\caption{Influence of \(\beta = 0\) on different types of loss errors.}
\begin{tabular}{lccc}
\hline
\textbf{Loss} & \textbf{Fidelity} & \textbf{Wasserstein} & \textbf{JSD} \\
\hline
NETSW  & 0.938 $\pm$ 0.02 & 0.945 $\pm$ 0.06 & 0.933 $\pm$ 0.01 \\
FLWDS  & 0.991 $\pm$ 0.06 & 0.902 $\pm$ 0.02 & 0.909 $\pm$ 0.04 \\
PRECSC & 0.988 $\pm$ 0.02 & 0.991 $\pm$ 0.04 & 0.903 $\pm$ 0.06 \\
PRECC  & 0.908 $\pm$ 0.07 & 0.915 $\pm$ 0.04 & 0.903 $\pm$ 0.02 \\
SOLS   & 0.968 $\pm$ 0.07 & 0.972 $\pm$ 0.05 & 0.973 $\pm$ 0.05 \\
SOLL   & 0.947 $\pm$ 0.01 & 0.955 $\pm$ 0.08 & 0.959 $\pm$ 0.10 \\
SOLSD  & 0.921 $\pm$ 0.07 & 0.945 $\pm$ 0.04 & 0.935 $\pm$ 0.01 \\
SOLLD  & 0.958 $\pm$ 0.03 & 0.947 $\pm$ 0.02 & 0.961 $\pm$ 0.05 \\
\hline
\end{tabular}
\end{table}

\begin{table}[htbp]
\centering
\caption{Influence of \(\beta = 3\) on different types of loss errors.}
\begin{tabular}{lccc}
\hline
\textbf{Loss} & \textbf{Fidelity} & \textbf{Wasserstein} & \textbf{JSD} \\
\hline
NETSW  & 0.743 $\pm$ 0.02 & 0.751 $\pm$ 0.06 & 0.739 $\pm$ 0.01 \\
FLWDS  & 0.805 $\pm$ 0.06 & 0.812 $\pm$ 0.02 & 0.818 $\pm$ 0.04 \\
PRECSC & 0.896 $\pm$ 0.02 & 0.904 $\pm$ 0.04 & 0.913 $\pm$ 0.06 \\
PRECC  & 0.715 $\pm$ 0.07 & 0.722 $\pm$ 0.04 & 0.714 $\pm$ 0.02 \\
SOLS   & 0.779 $\pm$ 0.07 & 0.785 $\pm$ 0.05 & 0.782 $\pm$ 0.05 \\
SOLL   & 0.756 $\pm$ 0.01 & 0.763 $\pm$ 0.08 & 0.770 $\pm$ 0.10 \\
SOLSD  & 0.841 $\pm$ 0.07 & 0.859 $\pm$ 0.04 & 0.848 $\pm$ 0.01 \\
SOLLD  & 0.872 $\pm$ 0.03 & 0.866 $\pm$ 0.02 & 0.881 $\pm$ 0.05 \\
\hline
\end{tabular}
\end{table}

\begin{table}[htbp]
\centering
\caption{Influence of \(\beta = 6\) on different types of loss errors.}
\begin{tabular}{lccc}
\hline
\textbf{Loss} & \textbf{Fidelity} & \textbf{Wasserstein} & \textbf{JSD} \\
\hline
NETSW  & \textbf{0.632} $\pm$ 0.02 & \textbf{0.644} $\pm$ 0.06 & \textbf{0.630} $\pm$ 0.01 \\
FLWDS  & \textbf{0.684} $\pm$ 0.06 & \textbf{0.700} $\pm$ 0.02 & \textbf{0.703} $\pm$ 0.04 \\
PRECSC & \textbf{0.779} $\pm$ 0.02 & \textbf{0.785} $\pm$ 0.04 & \textbf{0.799} $\pm$ 0.06 \\
PRECC  & \textbf{0.603} $\pm$ 0.07 & \textbf{0.607} $\pm$ 0.04 & \textbf{0.601} $\pm$ 0.02 \\
SOLS   & \textbf{0.655} $\pm$ 0.07 & \textbf{0.671} $\pm$ 0.05 & \textbf{0.666} $\pm$ 0.05 \\
SOLL   & \textbf{0.644} $\pm$ 0.01 & \textbf{0.652} $\pm$ 0.08 & \textbf{0.651} $\pm$ 0.10 \\
SOLSD  & \textbf{0.716} $\pm$ 0.07 & \textbf{0.740} $\pm$ 0.04 & \textbf{0.729} $\pm$ 0.01 \\
SOLLD  & \textbf{0.748} $\pm$ 0.03 & \textbf{0.737} $\pm$ 0.02 & \textbf{0.751} $\pm$ 0.05 \\
\hline
\end{tabular}
\end{table}

\newpage
To further perform the analytical ablation study of the results shown in Table 7, we need to evaluate the performance of different outputs (NETSW, FLWDS, PRECSC, PRECC, SOLS, SOLL, SOLSD, SOLLD) across three latent diameters (1, 5, and 7) using two evaluation models (ED and QED).

As the latent diameter increases, the capacity of the model to capture complex data structures also increases. For NETSW, a consistent high performance in both ED and QED across all latent diameters suggests that this parameter effectively leverages the additional capacity provided by larger latent spaces. The slight improvements in QED values from 0.947 at latent diameter 1 to 0.981 at latent diameter 7 indicate that NETSW can better capture the quantitative aspects of the data with increased dimensionality. FLWDS shows moderate performance improvements with increasing latent diameter. While it starts with lower ED and QED values at a latent diameter of 1, these values improve slightly at larger latent diameters. This trend indicates that FLWDS benefits from an increased latent space, enhancing its ability to represent more complex data patterns. However, its performance remains lower than NETSW, suggesting a limitation in its overall robustness.

The consistently poor performance of PRECC, with significantly negative ED and QED values across all latent diameters, highlights a fundamental issue with this parameter. The negative values indicate an inability to effectively model the data, and this inadequacy persists regardless of the latent space size. This suggests that increasing the latent diameter does not resolve the underlying problems in PRECC's data representation capabilities. SOLS, SOLSD, and SOLL exhibit robust performance similar to NETSW, maintaining high ED and QED values across different latent diameters. This consistency demonstrates their ability to effectively utilize the increased latent space to capture complex data structures. The stable or slightly improved performance metrics as the latent diameter increases reinforce their reliability and adaptability.

Increasing the latent diameter from 1 to 7 generally leads to enhanced model performance for robust parameters like NETSW, SOLS, and SOLSD. These parameters show that a larger latent space provides additional capacity to capture more intricate patterns and relationships in the data, which is reflected in the high and stable ED and QED values. Conversely, the persistent poor performance of PRECC across all latent diameters indicates that the problems with this parameter are intrinsic and not mitigated by increasing the latent space.

\begin{table}[htbp]
\centering
\caption{The influence of number of latent diameters on \(R^2\) values for QED and ED models.}
\begin{tabular}{lcc cc cc}
\hline
\textbf{Variable} & \multicolumn{2}{c}{\textbf{Latent Diameter = 1}} & \multicolumn{2}{c}{\textbf{Latent Diameter = 5}} & \multicolumn{2}{c}{\textbf{Latent Diameter = 7}} \\
\hline
 & \textbf{ED (\(R^2\))} & \textbf{QED (\(R^2\))} & \textbf{ED (\(R^2\))} & \textbf{QED (\(R^2\))} & \textbf{ED (\(R^2\))} & \textbf{QED (\(R^2\))} \\
\hline
NETSW  & 0.980 & 0.947 & 0.980 & 0.986 & 0.980 & 0.981 \\
FLWDS  & 0.802 & 0.872 & 0.802 & 0.895 & 0.802 & 0.890 \\
PRECSC &   -   &   -   &   -   &   -   &   -   &   -   \\
PRECC  & -17.909 & -15.9 & -17.909 & -12.9 & -17.909 & -14.89 \\
SOLS   & 0.960 & 0.966 & 0.960 & 0.973 & 0.960 & 0.962 \\
SOLL   & 0.945 & 0.950 & 0.945 & 0.965 & 0.945 & 0.957 \\
SOLSD  & 0.951 & 0.956 & 0.951 & 0.966 & 0.951 & 0.959 \\
SOLLD  & 0.857 & 0.881 & 0.857 & 0.898 & 0.857 & 0.891 \\
\hline
\end{tabular}
\end{table}

The Structural Similarity Index (SSIM) is a metric used to evaluate the similarity between two images, ranging from 0 to 1, where 1 signifies perfect similarity. This analysis aims to compare the SSIM values obtained from various configurations and training methodologies, specifically focusing on the Quantum Encoder-Decoder (QED) and its classical counterpart.

\begin{figure}[H]
    \centering
    \includegraphics[width=1\textwidth]{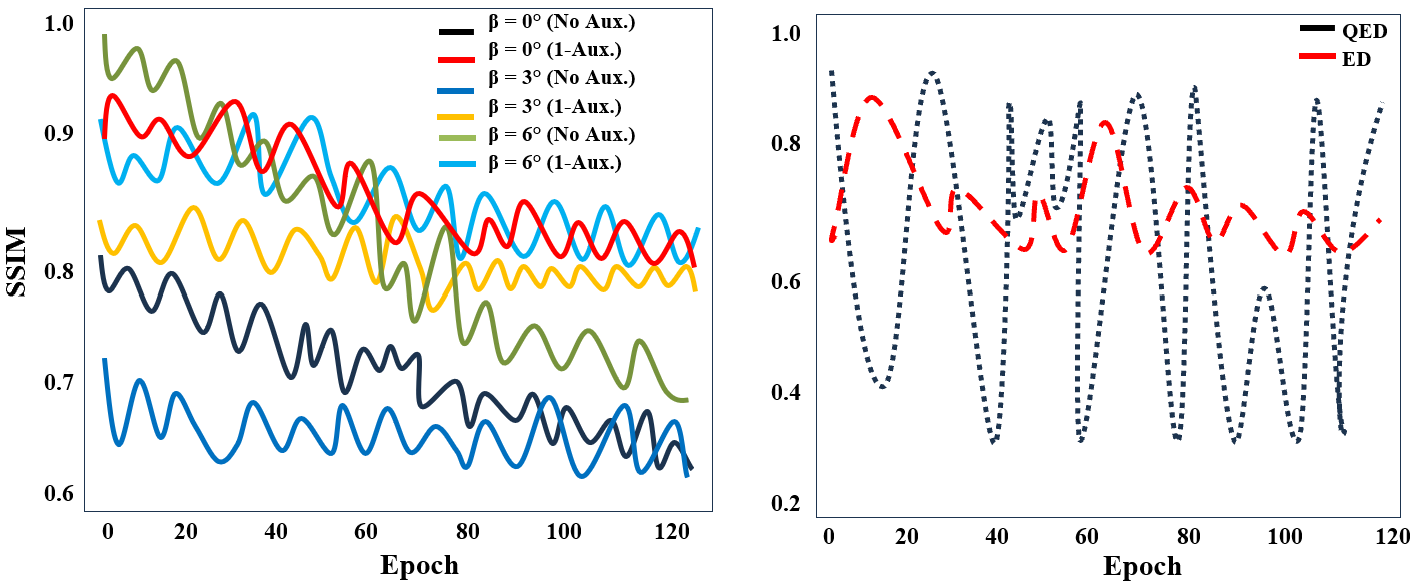}
    \caption{The effect of $\beta$ values on SSIM metric for QED and ED models.}
    \label{fig:qed}
\end{figure}

As can be seen in Figure 11, the influence of different $\beta$ values and the presence of auxiliary losses on SSIM over 120 epochs has been examined. Across the configurations tested, it is evident that auxiliary losses consistently improve SSIM values. This indicates that auxiliary losses are effective in enhancing image reconstruction quality by providing additional training signals that help the model learn more robust features. Comparing different $\beta$ values, it is clear that higher $\beta$ values yield better SSIM performance. Specifically, configurations with $\beta = 6^\circ$ consistently exhibit the highest SSIM values, indicating that greater rotational perturbations during training help the model generalize better to the reconstruction task. 

Among the tested values, $\beta = 6^\circ$ with auxiliary loss provides the best performance, demonstrating the combined effect of optimal rotation and auxiliary training strategies. Also, we compare the SSIM performance of QED and ED methods. The QED method, represented by the black curve, shows significant variability in SSIM values, ranging from 0.4 to 0.9. This variability suggests that while QED can occasionally achieve high SSIM, it is highly sensitive to noise and other perturbations, leading to unstable performance. In contrast, the ED method, represented by the red curve, maintains relatively stable SSIM values, mostly between 0.6 and 0.8. This stability indicates that ED is more robust to variations, providing more consistent image quality despite not reaching the highest SSIM values observed with QED.
\vspace{10pt}

\begin{figure}[H]
    \centering
    \includegraphics[width=1\textwidth]{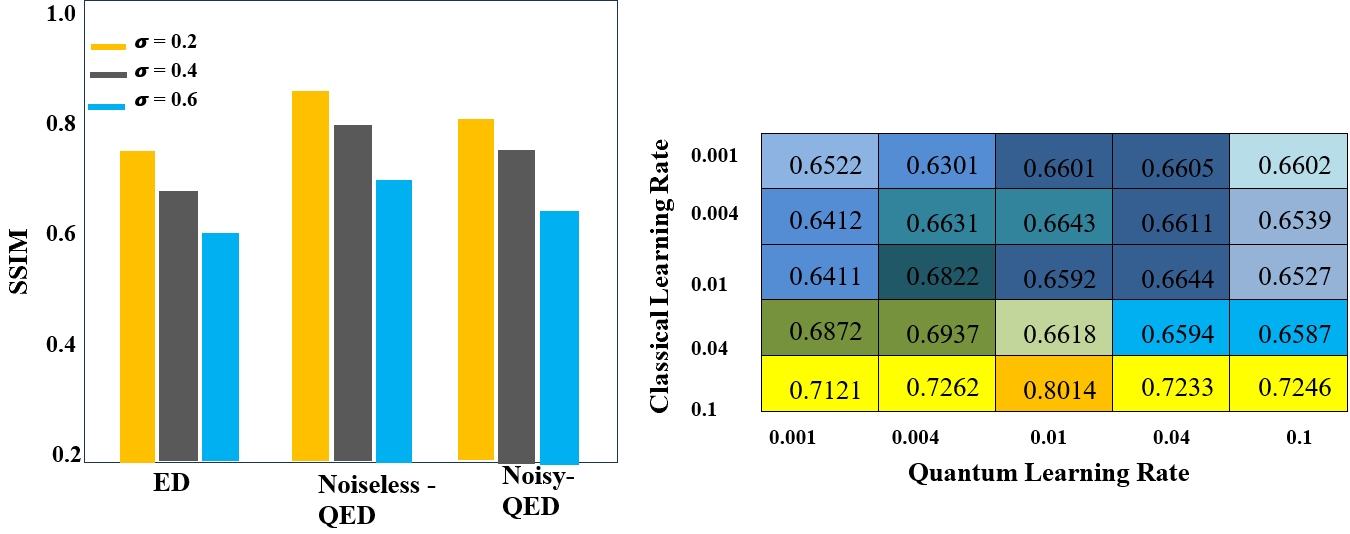}
    \caption{The effect of noise level and quantum/classical learning rate on SSIM metric.}
    \label{fig:qed}
\end{figure}

Further ablation study exploration ends up with the comparison across different noise levels, revealing that the SSIM values for the three denoising techniques (ED, Noiseless-QED, Noisy-QED) decrease as the noise level ($\sigma$) increases. Noiseless-QED consistently shows the highest SSIM values across all noise levels, indicating superior performance in denoising compared to ED and Noisy-QED. Noisy-QED, while slightly lower than Noiseless-QED, still outperforms the classical ED method, especially at lower noise levels.

From Figure 12, focusing on technique-specific performance, the classical denoising method (ED) achieves an SSIM of approximately 0.75 at $\sigma = 0.2$, dropping to around 0.65 at $\sigma = 0.4$, and further reducing to around 0.55 at $\sigma = 0.6$. This indicates that the performance of ED degrades significantly with increasing noise, showing less robustness to high noise levels. In contrast, Noiseless-QED (quantum-enhanced denoising in ideal conditions) achieves an SSIM close to 0.9 at $\sigma = 0.2$, around 0.85 at $\sigma = 0.4$, and remains robust at around 0.75 at $\sigma = 0.6$. This demonstrates that Noiseless-QED maintains high SSIM values even at higher noise levels, showcasing the effectiveness of quantum enhancement in ideal conditions. Similarly, Noisy-QED (quantum-enhanced denoising in realistic conditions) achieves an SSIM around 0.85 at $\sigma = 0.2$, dropping slightly to around 0.75 at $\sigma = 0.4$, and to around 0.7 at $\sigma = 0.6$. While the performance is slightly reduced compared to Noiseless-QED, Noisy-QED still outperforms ED, indicating the robustness of the quantum-enhanced method even in realistic noise conditions. Analytical insights reveal that the efficiency of quantum methods, leveraging the computational advantages of quantum algorithms, results in higher SSIM values and better image quality retention. Noisy-QED's performance in realistic conditions highlights its potential for real-world applications where noise is inevitable. Further optimization of quantum circuits and error mitigation techniques can potentially close the gap between Noisy-QED and Noiseless-QED, enhancing practical applicability.

According to Figure 12, the heatmap displays the SSIM values for different combinations of classical and quantum learning rates. A key observation is that as the quantum learning rate increases from 0.001 to 0.1, there is a general increase in SSIM values. This suggests that higher quantum learning rates improve the training efficiency and image reconstruction quality. Similarly, increasing the classical learning rate from 0.001 to 0.1 also shows an improvement in SSIM values, although the effect is more pronounced at specific quantum learning rates. The highest SSIM value (0.8014) is observed at a quantum learning rate of 0.01 and a classical learning rate of 0.1. This indicates that this combination is the most effective in balancing the training dynamics between the classical and quantum components of the model. Other high SSIM values (above 0.7) are concentrated around the same region, particularly with classical learning rates of 0.04 and 0.1 paired with quantum learning rates of 0.01, 0.04, and 0.1. This highlights a synergy between moderate to high learning rates in both domains.

At the lower end of the learning rates (0.001 for both quantum and classical), SSIM values are relatively lower, indicating underfitting. The model may not be learning effectively, resulting in poorer image reconstruction. At the highest classical learning rate of 0.1 paired with the lowest quantum learning rate of 0.001, the SSIM value is also lower (0.7121), suggesting potential instability or overfitting in the classical part while the quantum part lags in learning. The results underscore the importance of balanced learning dynamics between the classical and quantum components. Effective learning occurs when both components are able to optimize their respective parameters without significant lag from one side. The interaction effects between classical and quantum learning rates are evident, where specific combinations yield better performance. This emphasizes the need for careful tuning of hyperparameters in hybrid models to achieve optimal performance.

\section{Experiments}
We individually assess the accuracy of each component within the output vector using distinct evaluation metrics. For instance, we calculate Mean Absolute Error (MAE) and the coefficient of determination (R$^2$) at each point in the horizontal and vertical dimensions, then average these values across both dimensions. This process results in the summary statistics presented in Table 8 and Table 9, showcasing the performance of various baseline models across different output variables.

\begin{table}[htbp]
\centering
\caption{MAE [W/m$^2$] for baseline models.}
\begin{tabular}{lccccccc}
\toprule
\textbf{Variable} & \textbf{CNN} & \textbf{QCNN} & \textbf{MLP} & \textbf{QMLP} & \textbf{ED} & \textbf{QED} & \textbf{QME} \\ 
\midrule
Surface pressure [Pa]    & 18.85  & 14.65  & 13.36  & \textbf{9.456}  & 14.968 & 14.968 & \textbf{7.226} \\
Insolation [W/m$^2$]     & 8.598  & 4.138  & 5.224  & \textbf{2.541}  & 6.894  & 6.894  & \textbf{1.201} \\
Latent heat flux [W/m$^2$] & 3.364  & 1.122  & 2.684  & \textbf{0.364}  & 3.046  & 3.046  & \textbf{0.114} \\
Sensible heat flux [W/m$^2$]  & 37.83  & 25.90  & 34.33  & \textbf{21.91}  & 37.250 & 37.250 & \textbf{15.88} \\
Surface temperature [K]  & 10.83  & 6.54   & 7.971  & \textbf{3.871}  & 8.554  & 8.554  & \textbf{1.960} \\
Precipitation [mm/day]   & 13.15  & 8.02   & 10.30  & \textbf{5.130}  & 10.924 & 10.924 & \textbf{3.742} \\
Surface wind speed [m/s] & 5.817  & 2.673  & 4.533  & \textbf{0.993}  & 5.075  & 5.075  & \textbf{0.486} \\
Longwave radiation [W/m$^2$] & 5.679  & 2.879  & 4.806  & \textbf{1.403}  & 5.136  & 5.136  & \textbf{0.752} \\
\bottomrule
\end{tabular}
\end{table}

\begin{table}[htbp]
\centering
\caption{R$^2$ for baseline models.}
\begin{tabular}{lccccccc}
\toprule
\textbf{Variable} & \textbf{CNN} & \textbf{QCNN} & \textbf{MLP} & \textbf{QMLP} & \textbf{ED} & \textbf{QED} & \textbf{QME} \\ 
\midrule
Surface pressure [Pa]    & 0.944 & 0.962 & 0.983 & \textbf{0.995} & 0.980 & 0.986 & \textbf{0.997} \\
Insolation [W/m$^2$]     & 0.828 & 0.909 & 0.924 & \textbf{0.974} & 0.802 & 0.895 & \textbf{0.983} \\
Latent heat flux [W/m$^2$] & -     & -     & -     & -              & -     & -     & -                \\
Sensible heat flux [W/m$^2$]  & 0.077 & \textbf{0.115} & -38.69 & -22.59 & -17.909 & -12.9 & \textbf{0.091} \\
Surface temperature [K]  & 0.927 & 0.957 & 0.961 & \textbf{0.991} & 0.960 & 0.973 & \textbf{0.994} \\
Precipitation [mm/day]   & 0.916 & 0.968 & 0.948 & \textbf{0.988} & 0.945 & 0.965 & \textbf{0.992} \\
Surface wind speed [m/s] & 0.927 & 0.935 & 0.956 & \textbf{0.982} & 0.951 & 0.966 & \textbf{0.987} \\
Longwave radiation [W/m$^2$] & 0.813 & 0.887 & 0.866 & \textbf{0.929} & 0.857 & 0.898 & \textbf{0.942} \\
\bottomrule
\end{tabular}
\end{table}
\vspace{10pt}
The Table 10 compares the performance of different classical and quantum models based on their Root Mean Square Error (RMSE) values for various climate-related variables. The models analyzed are the Convolutional Neural Network (CNN), Quantum Convolutional Neural Network (QCNN), Multi-Layer Perceptron (MLP), Quantum Multi-Layer Perceptron (QMLP), Encoder-Decoder (ED), and Quantum Encoder-Decoder (QED).

Across all variables, the Quantum Multi-Layer Perceptron (QMLP) consistently achieves the lowest RMSE values, indicating its superior predictive accuracy. For example, in the case of NETSW (Net Shortwave Radiation) and FLWDS (Downwelling Longwave Radiation), QMLP significantly outperforms other models, demonstrating its strength in capturing complex relationships in the data.

The Quantum Convolutional Neural Network (QCNN) also performs well, generally better than its classical counterpart, the CNN. This trend is evident in variables such as PRECSC (Precipitation Convective Scheme) and PRECC (Precipitation), where QCNN shows notably lower RMSE values compared to CNN. This suggests that incorporating quantum computing elements into neural networks can enhance their ability to model intricate patterns in climate data.

Classical models like CNN and MLP show higher RMSE values across most variables, indicating lower predictive accuracy compared to their quantum-enhanced versions. This disparity highlights the potential limitations of classical models in handling the complexity of climate-related data. The Encoder-Decoder (ED) and Quantum Encoder-Decoder (QED) models exhibit moderate performance, with QED typically outperforming ED. This further supports the notion that quantum models can provide an edge in predictive tasks over their classical equivalents.

In summary, the analysis highlights the superiority of quantum models, particularly the QMLP, in achieving lower RMSE values across a range of climate variables. The consistent performance of QCNN also underscores the benefits of quantum computing in enhancing neural network capabilities. Needless to mention that the QME (which is the ensemble analysis of the quantum models) achieved the highest accuracy among the rest of existing models. These findings suggest that quantum-enhanced models hold significant promise for improving predictive accuracy in climate analysis and related fields.

\begin{table}[htbp]
\centering
\caption{The RMSE results for different classical and quantum models.}
\begin{tabular}{lcccccccc}
\hline
\textbf{Variable} & \textbf{CNN} & \textbf{QCNN} & \textbf{MLP} & \textbf{QMLP} & \textbf{ED} & \textbf{QED} & \textbf{QME} \\
\hline
NETSW  & 36.91 & 24.54 & 26.71 & \textbf{18.476} & 28.537 & 21.968 & \textbf{14.564} \\
FLWDS  & 10.86 & 6.138 & 6.969 & \textbf{3.533}  & 9.070  & 6.732  & \textbf{2.133}  \\
PRECSC & 6.001 & 3.102 & 4.734 & \textbf{2.314}  & 5.078  & 4.046  & \textbf{1.092}  \\
PRECC  & 85.31 & 75.90 & 72.88 & \textbf{59.90}  & 76.682 & 67.296 & \textbf{36.82}  \\
SOLS   & 22.92 & 16.54 & 17.40 & \textbf{13.87}  & 17.999 & 15.524 & \textbf{8.874}  \\
SOLL   & 27.25 & 18.02 & 21.95 & \textbf{15.13}  & 22.540 & 17.864 & \textbf{11.75}  \\
SOLSD  & 12.13 & 7.679 & 9.420 & \textbf{4.998}  & 9.917  & 6.575  & \textbf{2.762}  \\
SOLLD  & 12.10 & 7.842 & 10.12 & \textbf{6.403}  & 10.417 & 7.036  & \textbf{3.971}  \\
\hline
\end{tabular}
\end{table}
\vspace{10pt}

When analyzing the Mean Absolute Error (MAE) and R-squared (R$^2$) values, QME emerges as the most effective model across almost all variables. For instance, in the NETSW variable, QME achieves the lowest MAE (7.226 W/m$^2$) and the highest R$^2$ (0.997), indicating its superior predictive accuracy and reliability. The second-best model, QMLP, also performs well with an MAE of 9.456 W/m$^2$ and an R$^2$ of 0.995, but it still falls short of the ensemble approach of QME. The similar trend is observed for FLWDS as QME hits the best prediction accuracy. For variables such as PRECSC and PRECC, where the classical models struggle to achieve high accuracy, quantum models and ensemble show marked improvements. The superiority of QME and QMLP is further validated by the SOLS, SOLL, SOLSD, and SOLLD variables. 

For the sake of further confirming these results, the R$^2$ distribution maps for FLWDS and PRECSC are visualized in Figure 13 accordingly as classical models like CNN, MLP, and ED have lower R$^2$ values across the globe, indicating less accurate predictions. In contrast, quantum models such as QCNN, QED, and QMLP exhibit higher R$^2$ values, suggesting better performance and higher accuracy. This is visually represented by the extensive areas in red and orange shades in the R$^2$ distribution maps, indicating higher correlation and predictive power.

As can be seen in Figure 14, the $R^2$ regions with lower values are identified for output variable distributions such as NETSW, FLWDS, and PRECSC. Accordingly, it can be clearly observed that the surface areas and number of regions with lower $R^2$ have decreased once we tend to utilize QME and QMLP instead of QCNN, QED, and the rest of classical models. In this regard, QME and QMLP outperform the existing quantum ML models along with their classical ML counterparts.

\begin{figure}[H]
    \centering
    \includegraphics[width=1\textwidth]{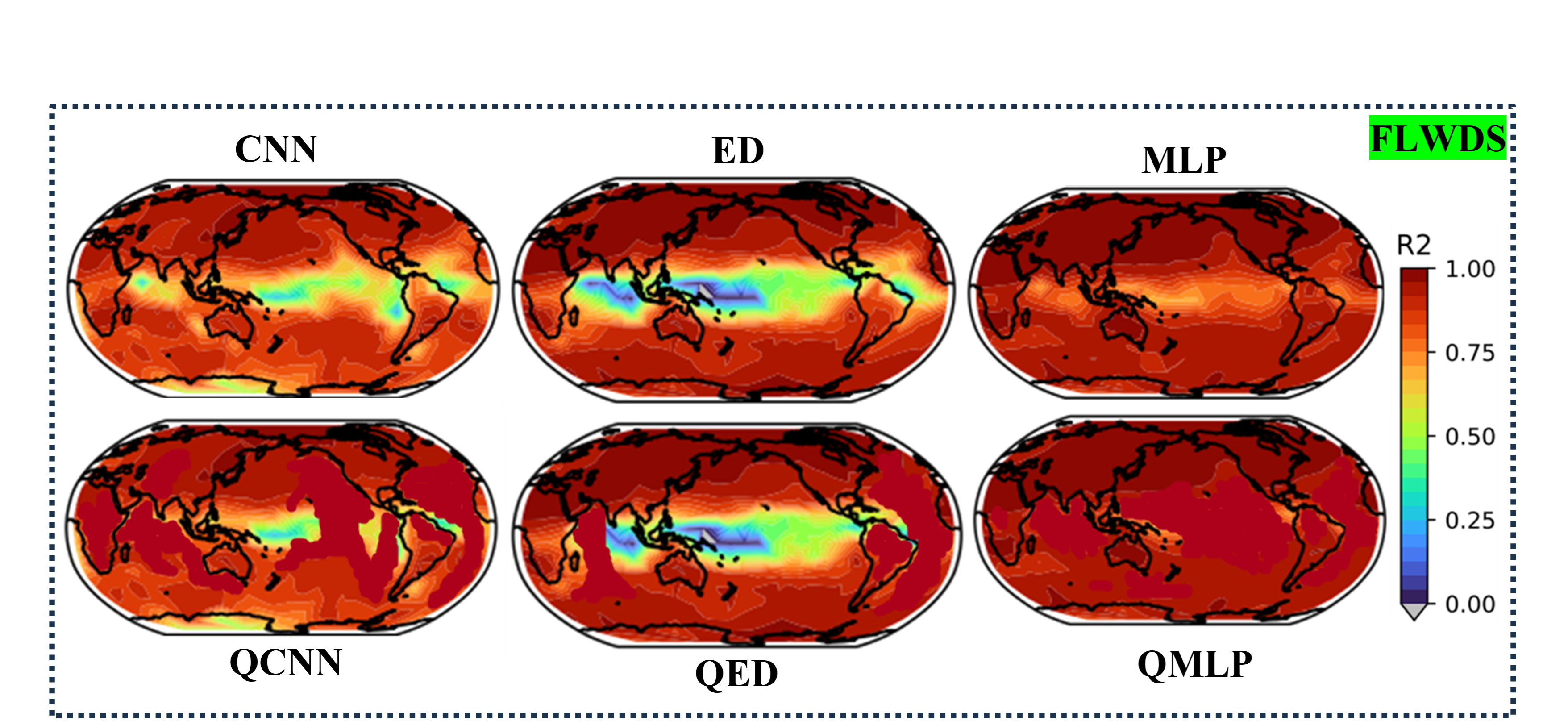}
    \centering
    \includegraphics[width=1\textwidth]{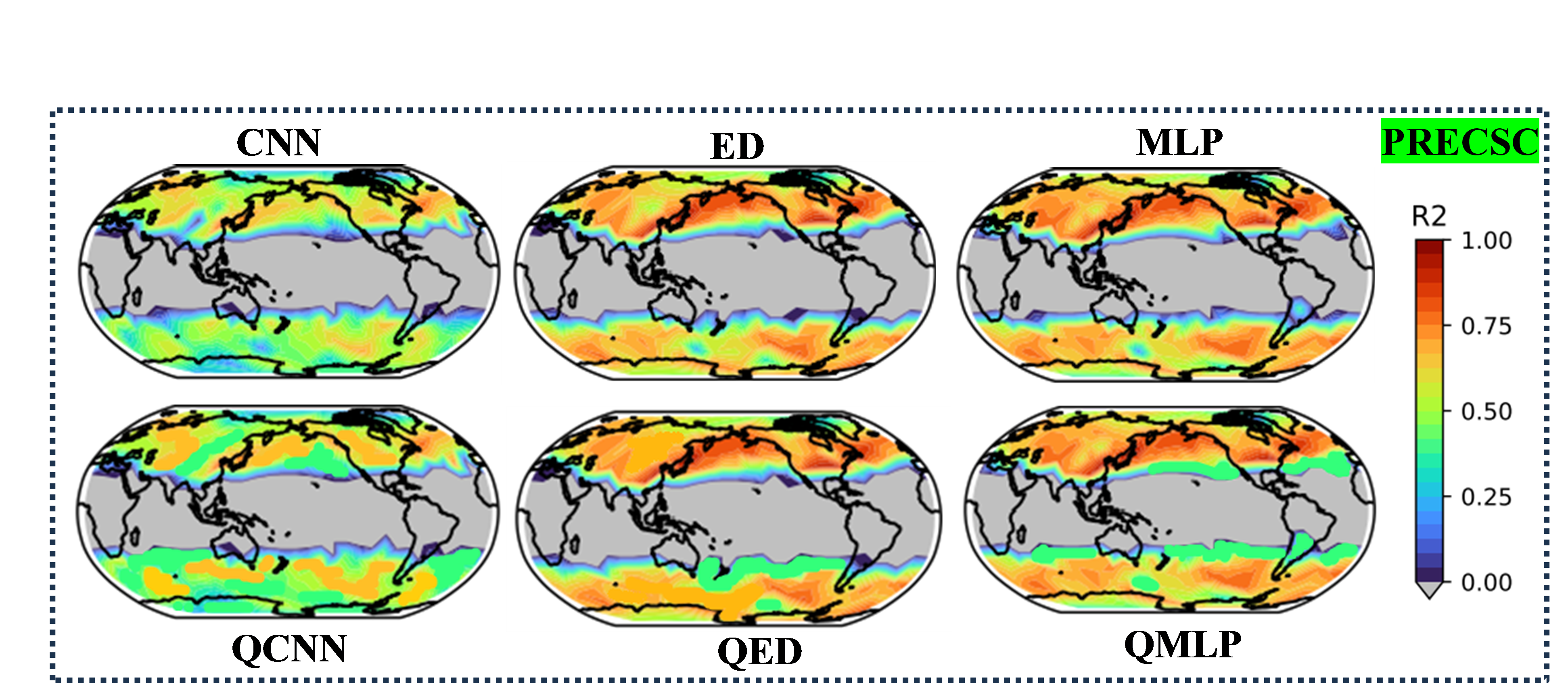}
    \caption{Quantum baseline model predictions for FLWDS and PRECSC.}
    \label{fig:qed}
\end{figure}

\begin{figure}[H]
    \centering
    \includegraphics[width=1\textwidth]{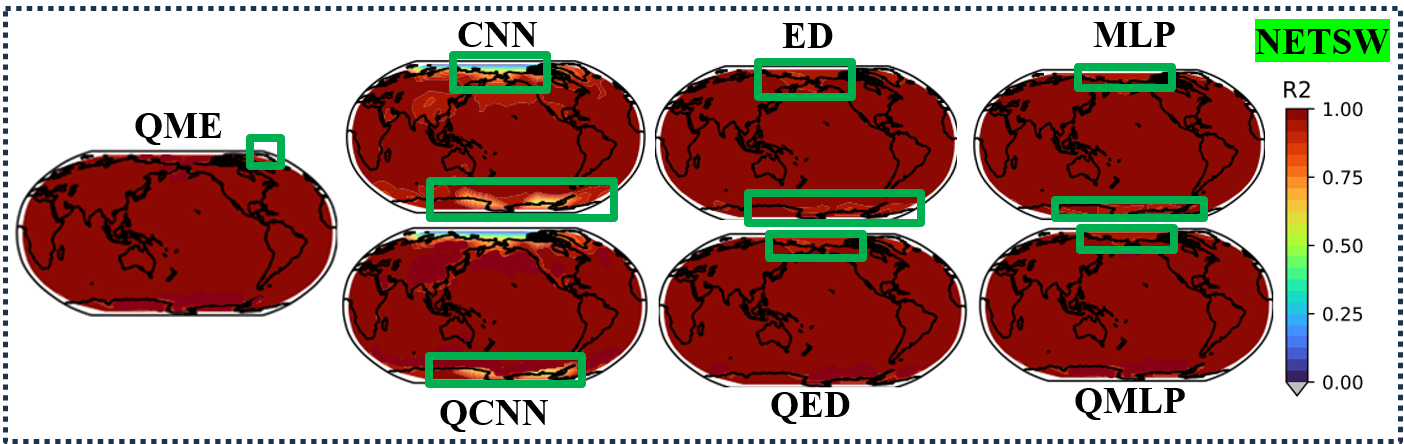}
    \centering
    \includegraphics[width=1\textwidth]{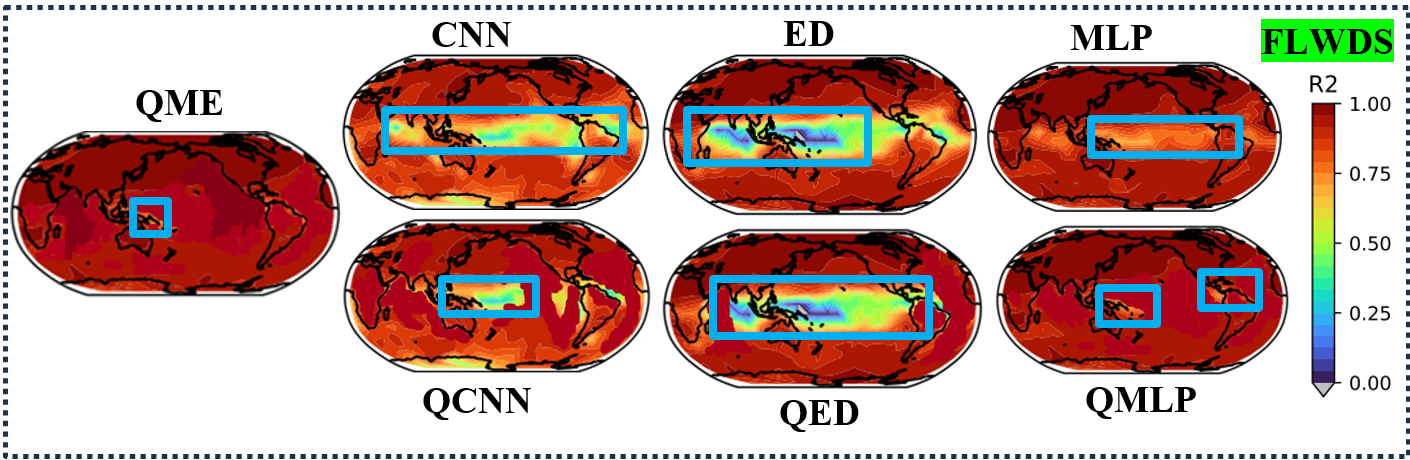}
    \centering
    \includegraphics[width=1\textwidth]{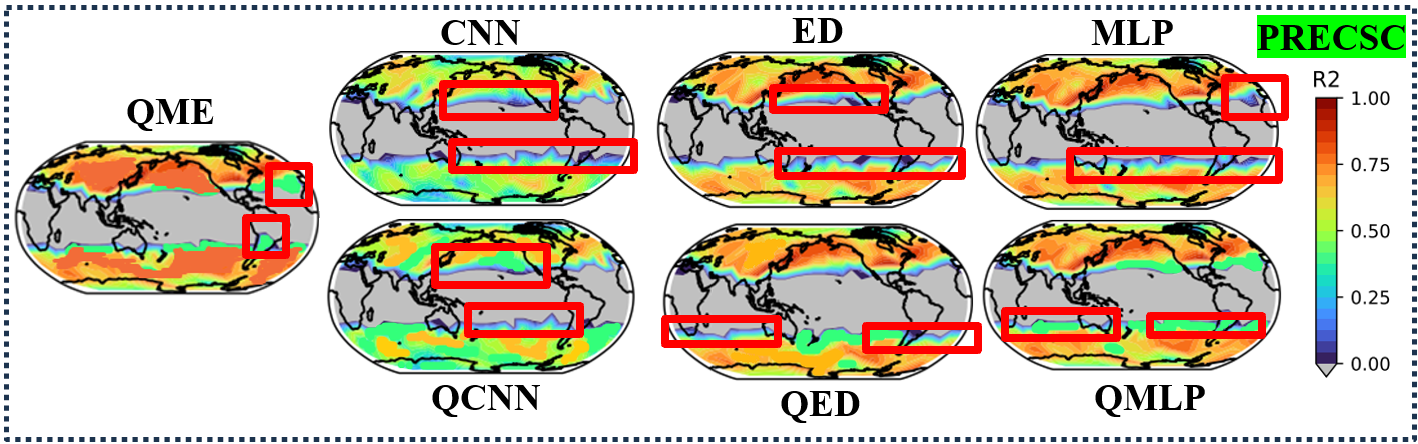}
    \caption{Quantum baseline model predictions for NETSW, FLWDS, and PRECSC.}
    \label{fig:qed}
\end{figure}

\section{Conclusion}
The adoption of quantum-enhanced machine learning models, such as QCNN, QMLP, and QED, represents a significant step forward in Earth and climate modeling. These models outperform traditional methods by offering better prediction accuracy and efficiency, especially in handling complex atmospheric phenomena. The meta-ensemble approach further boosts predictive performance, suggesting that quantum models not only excel individually but also synergize effectively. Despite their promise, quantum models face challenges, particularly in terms of the large computational resources required for training. Nonetheless, advancements in quantum hardware and algorithms could unlock the full potential of these models, making them indispensable for more accurate climate predictions. In conclusion, quantum-enhanced climate models present a powerful tool for improving the accuracy and efficiency of climate simulations, offering valuable insights for both scientific research and policy-making. Continued innovation in this field will be key to addressing the pressing challenges posed by climate change.

\section{Data and Code Availability} The code and data developed in this study can be found in our GitHub repository at \textbf{\href{https://github.com/adibgpt/QESM}{QESM GitHub}}.

\end{document}